\documentclass[useAMS,usenatbib,usegraphicx]{mn2e}
 
\title[Galaxy Formation as a  Cosmological Probe]{Galaxy Formation as a Cosmological Tool. I: The Galaxy Merger History as a Measure of Cosmological Parameters}
\author[C.J. Conselice et al.]{Christopher J. Conselice$^{1}$\thanks{E-mail:
conselice@nottingham.ac.uk}, Asa F.L. Bluck$^{2}$, Alice Mortlock$^{1,3}$, David Palamara$^{4,5}$, \newauthor Andrew J. Benson$^{6}$\\
$^{1}$University of Nottingham, School of Physics \& Astronomy, Nottingham, NG7 2RD UK \\
$^{2}$University of Victoria, Department of Physics and Astronomy, Victoria, British Columbia, V8P 1A1, Canada \\
$^{3}$Royal Observatory Edinburgh \\
$^{4}$School of Physics, Monash University, Clayton, Victoria 3800, Australia \\
$^{5}$Monash Center for Astrophysics (MoCA), Monash University, Clayton, Victoria 3800, Australia \\\
$^{6}$Carnegie Observatories, 813 Santa Barbara Street, Pasadena, CA 91101, USA}

\def\solm{M$_{\odot}\,$}

\def\solm{M$_{\odot}\,$}

\def\om{$\Omega_{\rm m}\,$}
\def\ol{$\Omega_{\Lambda}\,$}
\def\mhalo{M$_{\rm halo}\,$}

\def\casgm20{CAS-G-M$_{20}\,$}
\def\m20{M$_{20}\,$}
\begin{document}

\date{Accepted ; Received ; in original form}
\pagerange{\pageref{firstpage}--\pageref{lastpage}} \pubyear{2002}

\maketitle

\label{firstpage}

\begin{abstract}

As galaxy formation and evolution over long cosmic
time-scales depends to a large degree on the structure of the universe,
the assembly history of galaxies is potentially a powerful approach for
learning about the universe itself.  In this paper we examine the merger
history of dark matter halos based on the Extended Press-Schechter 
formalism as a function of cosmological parameters, redshift and halo mass.  
We calculate how major halo mergers are influenced by changes 
in the cosmological values of $\Omega_{\rm m}$, $\Omega_{\Lambda}$, 
$\sigma_{8}$, the dark matter particle temperature 
(warm vs. cold dark matter), and the value of a 
constant and evolving equation of state parameter $w(z)$.
We find that the merger fraction at a given halo mass varies by up 
to a factor of three for halos forming under the assumption of Cold 
Dark Matter, within different 
underling cosmological parameters.   We find that the current measurements
of the merger history, as measured through observed galaxy pairs as well
as through structure, are in agreement with the concordance cosmology
with the current best fit giving 
$1 - \Omega_{\rm m} = \Omega_{\rm \Lambda} = 0.84^{+0.16}_{-0.17}$.  
To obtain a
more accurate constraint competitive with recently measured cosmological
parameters from Planck and WMAP requires a measured merger accuracy of 
$\delta f_{\rm m} \sim 0.01$, implying surveys with an accurately measured 
merger history over 2 - 20 deg$^{2}$, which will be feasible with the 
next generation of imaging and spectroscopic surveys such as Euclid and 
LSST. 

\end{abstract}


\section{Introduction}

One of the major goals in science is determining the past
history and future evolution of the universe.  The determination of
this in a quantitative way 
has a long history, starting with the work of Hubble (1929) who determined
that the universe was expanding based on radial velocity and distance
measurements of galaxies.
This has continued using various approaches, including the use of 
Cosmic Microwave Background (CMB) measurements, with the most
recent work using e.g., WMAP, Planck and BICEP2 (e.g., Komatsu et al. 2011; 
Ade et al. 2013, 2014).  Currently, the use
of the CMB and type Ia supernova are the most common and
influential methods for measuring
cosmological parameters (e.g., Kessler et al. 2009), along with baryonic 
acoustic
oscillations and clustering measurements (e.g., Eisenstein et al. 2005;
Blake et al. 2011).

One of the dominant features of the current popular cosmological model
is that the universe's energy budget is perhaps dominated by a cosmological
constant - the so-called 
Dark Energy.    The major evidence for this Dark Energy is largely based on 
observations of the luminosities of supernova at various redshifts 
(e.g., Riess et al. 1998; Perlmutter et al. 1999). Other
evidence for Dark Energy comes from baryonic acoustic oscillations 
(e.g,. Eisenstein et al. 2005), and fluctuations in the cosmic background
radiation (e.g., Komatsu et al. 2011).
  The major result of this is that the universe appears to be
accelerating since $z < 1$, and perhaps undergoes a deceleration
phase at higher redshifts (Reiss et al. 2004). 

What is currently lacking within this cosmological paradigm is
physical evidence for the existence of Dark Energy, which in principle
can significantly
change the evolution of the constituents of the universe, of which galaxies
are the fundamental component.  The basic idea is that
if the universe is undergoing an acceleration phase, then 
the rate of structure formation
will decline with time, halting the growth of massive structures, such as
galaxy clusters (e.g., Allen et al. 2004; Vikhlinin et al. 2009).  
In fact, observations of the number densities of galaxy clusters can
be used as an alternative method for constraining Dark Energy
properties (e.g., Vikhlinin et al. 2009).  

Recently, with a basic but firm understanding of galaxy formation and evolution
it is now possible to go beyond observations of clusters, supernova, and
the cosmic background radiation to use 
galaxies themselves as a new probe of cosmology.   We explore in this
paper how cosmological properties 
affect the formation of galaxies throughout their history based
on examining the formation histories of dark matter halos.  This lets us
examine both how galaxy formation can be used as a probe of
cosmology, and how cosmology affects the formation of galaxies.  In this sense
the formation history of galaxies in the universe is potentially 
another probe of the
energy and kinematics of the universe.  

The use of galaxies for cosmology is not a new idea, and early attempts
to measure cosmological properties, largely the measurement of the
Hubble constant, relied on luminosities and
properties of stars and globular clusters in external galaxies.  
Several cosmological tests were also
proposed in the 1920-1930s that used the angular sizes, counts,
and surface brightness evolution of galaxies (e.g., Tolman 1930; 
Sandage 1988).  However, 
these approaches were largely abandoned once it was realized that galaxies
evolve significantly through time, and that the properties of nearby 
galaxies are not necessarily the same as  more distant galaxies. 

Since we are now becoming confident in the measurements of galaxy properties, 
and how at least massive galaxies evolve and form over time, especially since
$z \sim 3$ (e.g., Bluck et al. 2012; Conselice et al. 2013; Muzzin et al. 2013) we are in a 
position to reevaluate whether galaxy properties, and their evolution, 
can be used to determine features of the Universe.  While early cosmological
investigations were based on measuring the Hubble constant through the
distance-velocity relation, and later through trying to measure
the value of the deceleration parameter, $q_{0}$, new approaches
using the evolution of galaxy properties can potentially be used to derive
features of the dominant cosmological paradigm.  

We specifically investigate in this paper whether the evolution of galaxies 
is consistent with the currently accepted ideas concerning a 
$\Lambda$-dominated universe with a transition from deceleration to 
acceleration occurring sometime around $z \sim 0.7$ (Turner \& Riess 2002).  
The fact that the universe
may transition from one which is decelerating to one which is accelerating
may have a profound impact on the formation history of galaxies
which would otherwise be different in a universe with a different
cosmology.    The idea behind this paper is that the history of galaxy
assembly is driven by cosmological parameters, and one of the ways this can
be seen is through the merger history.  

In this first paper of this series we investigate this problem by comparing 
models of halo and galaxy formation, which vary as a function of cosmology 
and dark matter properties,
to the observed galaxy merger history.    We investigate the possibility that 
comparisons between halo merger histories and observed galaxy mergers 
can be used as an independent new
measurement of cosmological properties, and give us some indication of
what would be necessary to use these features to constrain cosmology
more directly using future telescopes and space missions.  To do this
we also discuss how comparing observations of galaxies to dark matter
halos is perhaps a better approach for understanding bulk galaxy formation
than to rely on the simulations of galaxy formation themselves.

  We furthermore show how comparisons between the galaxy merger history
predicted in CDM simulations differ depending on the underlying dark matter.
We demonstrate that the
temperature of the dark matter particle also can have a fundamental
influence on the predicted galaxy merger history.  We further conclude how comparing
halo merger models to observed galaxy mergers can reveal clues to
the selection of observed mergers and how, and whether, there is a 
self-consistent observational and cosmological based theoretical 
picture for the formation history of galaxies.

This paper is organized as follows: \S 2 gives a description of the models
we use in this paper, \S 3 includes a discussion of the results of an analysis
of halo merger histories. \S 4 is a description of the comparison between
galaxy merger history predictions, the actual merger history, and
consists of our main analysis, \S 5 
is a discussion of the implications for these results \S 6 is a discussion
of our results and \S 7 is our summary.
We refer to a cosmology with  \om = 0.3, \ol = 0.7; $\sigma_{8}$ = 0.9;
H$_{0} = 70$ km s$^{-1}$ Mpc$^{-3}$ as the concordance cosmology.

\section{Dark Matter Halo Models}

\subsection{Formalism}

In the dominant theory for galaxy formation, based on a $\Lambda$CDM
cosmology, galaxies assemble by merging with one another over time (e.g.,
White \& Rees 1978).  The basis for this merging is the dark matter 
assembly history, and how dark matter halo masses grow by merging 
with one another.    Using 
Newtonian dynamics plus
a simple expanding universe model it is now possible
to predict the total halo mass functions of galaxies across a large range in
mass to within 5\%, comparing different computer simulation results.  
With the small discrepancies based on the differences
 between methods of the various calculations rather than fundamental
physics (e.g., Reed et al. 2007). In particular, different group
finding algorithms are largely the cause of the small differences in
masses, rather than fundamental physics (e.g., Knebe et al. 2013).

Predictions for how structure assembles is the backbone of any theory of
galaxy formation.  Since galaxies are believed to form at the cores of dark 
matter halos,
then the formation of galaxies should follow in some way how the
dark matter assembles.  Dark matter halos and large scale structure
are created through these halos hierarchically.  This process
can be predicted based on the basic physics of gravitational 
collapse of matter in an expanding universe, and
its later evolution, and therefore does not involve uncertain 
baryonic physics.  The details of how dark matter halos assemble 
is now predicted in detailed N-body and semi-analytical simulations
(e.g., Fakhouri \& Ma 2008).  These simulations essentially predict when
two existing dark matter halos merge together to form a large halo
within the standard $\Lambda$CDM cosmologies assumed.

It is fairly straightforward to use simulations of structure formation
to predict how dark matter halos with descendant masses between
10$^{12}$ \solm $<$ M$_{\rm halo} < 10^{15}$ \solm assemble with 
time (e.g., Fakhouri \&
Ma 2008).  We investigate in this paper what various models predict for halo
mergers.  Our primary method is to use a generalized code for calculating
dark matter halo mergers within a given cosmology.  To do this
we calculate the merger history for dark matter halos through using the
`growl' algorithm by Hamilton (2001) using a power-spectrum
calculated by Eisenstein \& Hu (1999).   Using this numerical
formalism it is possible to determine the assembly history of dark
matter halos using basic gravitational collapse physics.

To calculate this we use the results of Hamilton (2001), and  a 
modified form of the {\em growl} code to compute the linear growth factor 

\begin{equation}
g = \frac{D}{a}
\end{equation}

\noindent for structure in the universe as a function of time,
where $D$ is the amplitude of the growth mode, and $a$ is the scale factor.
The linear growth rate, 

\begin{equation}
f = \frac{\rm d ln D}{{\rm d ln}\, a}
\end{equation}

\noindent is the derivative of $g$, and relates to peculiar motions within 
the universe.     Using
a Friedmannn-Robertson-Walker (FRW) universe, the growth factor $g$
can then be written as 

\begin{equation}
{\rm g(\Omega_{m},\Omega_{\Lambda})} = \frac{D}{a} = \frac{5 \times \Omega_{\rm m}}{2}\int_{0}^{1} \frac{{\rm d}a}{a^{3}H(a)^{3}}
\end{equation}

\noindent where $a$ is the scale-factor normalized to unity, and $H(a)$ 
is the Hubble parameter normalized
to unity when $a = 1$, where

\begin{equation}
H(a) = (\Omega_{\rm m} a^{-3} + \Omega_{\rm k} a^{-2} + \Omega_{\Lambda})^{1/2}.$$
\end{equation}
  
\noindent The value of the growth factor and the Hubble constant H($a$) 
depend upon the value of cosmological parameters.  The linear growth 
rate $f$ can then be written as 

\begin{equation}
f(\Omega_{\rm m}, \Omega_{\Lambda}) = -1 - \frac{\Omega_{\rm m}}{2} + \Omega_{\Lambda} + \frac{5 \Omega_{\rm m}}{2g}
\end{equation}

\noindent Analytical solutions to the above growth rate are presented in detail in
Hamilton (2001) for different cosmological parameter ratios. The {\em growl} 
code then implements these fitting formula for various 
scenarios to predict what the
growth factor is during the history of the universe, as a function
of cosmology and time.  

The power spectrum used within this code originates from 
Eisenstein \& Hu (1999), who calculate fitting formula for the 
matter transfer function as a function of
wavenumber, time, the massive neutrino density, number of neutrino species, the
Hubble parameter today, the cosmological constant, baryon density and 
the spatial curvature.  Mergers occur via the set excursion methodology
from Press \& Schechter (1974) but using the extended formalism.  
As a result, we measure the halo merger history as a function of the 
mass ratio  of the halo mergers:

\begin{equation}
\eta = \frac{M_{2} - M_{1}}{M_{2}},$$
\end{equation}

\noindent where $M_{2}$ is the sum of the halo masses of the two merger
components (or the resulting halo mass) and $M_{1}$ is the halo mass 
of the more massive progenitor.

We use these models with a variety of different cosmological parameters
to investigate how the halo merger history varies with cosmology.  
The cosmological parameters that we vary are: the matter 
density $\Omega_{\rm m}$, the dark energy density $\Omega_{\Lambda}$, 
the neutrino density $\Omega_{\mu}$, the Hubble constant, $H_{0}$, 
the baryonic density $\Omega_{\rm B}$, the temperature of the CMB 
T$_{\rm CMB}$, the number of neutrino species, N$_{\nu}$, value 
of $\sigma_{8}$, and the spectral index $n$. 
We define the cosmology henceforth as the quantity 

\begin{equation}
\Theta = \left({\Omega_{\rm m}, \Omega_{\Lambda}, \Omega_{\mu}, H_{0}, \Omega_{\rm B}, T_{\rm CMB}, N_{\nu}, \sigma_{8}, n}\right)$$  
\end{equation}

\noindent Our method uses an altered version of the publicly 
available {\em growl} code.  We measure the halo merger history through a 
particular type of `major'
halo merger. These cosmological based halo mergers are designed to match 
as much as possible
the merger criteria used to find mergers occurring in actual galaxies (\S 2.2).
We use these halo mergers as our primary method for comparing with 
observable galaxy mergers, as we show in \S 4 that the predicted galaxy 
merger history
is currently too uncertain to be used to compare with real galaxies to derive
cosmology, but that the halo mergers are known accurately enough to make
this comparison.

\subsection{Merger Fraction Calculation and Time-Scales} 

In this paper we only discuss mergers which are major for both the 
observational data and the theoretical results.
The criteria for finding a major merger is that a halo
at a given redshift must have had a merger with another halo of mass
1:4 or less within the past $\sim 0.4$ Gyr.  This is the same 
criteria we use to find mergers in galaxies based on pairs separated by 30 kpc 
(e.g., Bluck et al. 2012), and
when using the structural CAS system (e.g., Conselice et al. 2009; Conselice
2014). This
also matches well the mass ratio and time-scales for mergers for the CAS
parameters (e.g., Lotz et al. 2011).   We later discuss how these merger
fractions would change if using a different value of the merger time-scale 
and after matching halo vs. stellar mass ratio mergers.

We use a merger time-scale of 0.4 Gyr throughout this paper, as this is the 
average time-scale in which we are sensitive to within the observations  
based on both N-body models (Conselice 2006b; Lotz et al. 2008), and when 
examining the empirically measured merger time-scales (Conselice 2009).  Below
we discuss in detail the reasoning behind this, and the uncertainties 
associated with using a fixed time-scale.

First, it is clear that numerical models of massive galaxy mergers give an
average time-scale of $\sim 0.4$ Gyr.  For example, Conselice (2006b) and
later Lotz et al. (2008b, 2010a,b) investigate the location of 
different phases of various types of mergers in morphological parameter 
space.   They use these models to calculate the time-scales for how long these 
simulated galaxies appear as a `merger' ,
based on where they fall in these non-parametric structural spaces. While
Conselice (2006b) used only dark matter simulations, the Lotz et al.  
studies investigate the stellar distribution and how dust, viewing angle, 
orbital parameters, gas  properties, SN  feedback and total mass alter the 
merger time-scale (e.g., Lotz et al. 2010b; Moreno et al. 2013).
It is found in these papers that very few properties beyond mass ratio and 
gas mass fraction affect the derived merger time-scales.

These simulations show that mergers are identified within both CAS at the 
first pass of the
merger, as well as when the systems finally merge together to form a remnant 
(Lotz et al. 2008). However, merging galaxies are
not found in the merger area of the non-parametric structural parameters 
for the entire merger, as was found by Conselice (2006b).
This however allows the time-scales for structural mergers to be calculated. 
Lotz et al. (2008, 2010a) find that the asymmetry time-scales for gas-rich 
major mergers are 0.2-0.6 Gyr and 0.06 Gyr for minor mergers (Lotz et al. 
2010a).   While the individual time-scale for a pair of galaxies within a 
dark matter 
halo to merge will vary, based on the variety of models the average is 
0.4 Gyr and we use this throughout this paper as our measured merger 
time-scale.

An issue that we have to address is that these time-scales are for
gas rich mergers, and would not necessarily apply for gas poor 
or dry mergers. However,
at the redshifts we are investigating here, nearly all galaxy mergers will
have some gas, as pure dry mergers are relatively rare (e.g., Lin
et al. 2008; Conselice et al. 2009; De Propris et al. 2010).  Furthermore,
the massive galaxies that we examine in this paper have gas fractions 
which are on average $\sim 10$\% at $z = 1-3$, with little variation 
(e.g., Mannucci et al. 2009; Conselice et al. 2013).  Therefore it is 
unlikely that the computed merger time-scales 
differ due to a lack or over abundance of gas.   We however do calculate how 
our merger fractions would change if using a different time-scale. In summary
if our observed mergers have a longer/shorter time-scale then the resulting 
comparisons for halo mergers would be higher/lower. 

We utilize these numerical models
to determine the fraction of halos which have merged within our given
time-scale. For our purposes we determine the merger fraction of
halos, which we denote as $f_{\rm halo}$, at a variety of redshifts.  

\begin{equation}
$$f_{\rm halo} = \frac{N_{\rm merger}(M_{\rm halo}, \eta, z, \tau, \Theta)}{N_{\rm tot}(M_{\rm halo}, \eta, z, \tau, \Theta)}$$
\end{equation}

\noindent Later we investigate how the merger history of halos, and the value
of $f_{\rm halo}$ changes as a function of different values of the
equation of state parameter $\omega$, and for a varying $\omega(z)$
as a function of redshift.   These calculations originate from the 
{\em GALACTICUS} code of Benson (2012).  Within our calculations of the
merger fraction we also take into account double mergers whereby a halo
or galaxy undergoes more than one merger in a given time-scale.
The merger fraction includes the total number of mergers the examined
population (selected by mass in this case) undergoes divided by the total
number of galaxies in that selection.  Therefore if a single galaxy/halo
undergoes more than a single merger it is accounted for explicitly.

{\em GALACTICUS} is a semi-analytic model which  is easily adapted to 
differing physical and initial conditions.  We utilized the same frame-work 
presented in the Hamilton (2001) structure formation model, but through using
an equation of state parameter, $\omega$, as well as through the
use of an evolving form as a function of redshift ($\omega(z)$).

We also use the results from the Millennium simulations (Springel
et al. 2005) for both the merger history of galaxies which was discussed in 
depth in  Bertone \& Conselice (2009), as well as the merger history of 
halos.  We furthermore compare with Warm Dark Matter semi-analytical models 
from Menci et al. (2012) to test how different  dark matter particle 
temperatures can affect the galaxy merger history. Finally,
we also compare with abundance matched merger histories from
Hopkins et al. (2010).


\subsection{Data sources}

One of the major goals of this study is to compare the observed galaxy
merger history with predictions from simulations for the halo and predicted
galaxy merger history.  As such, the data we use for this comparison 
are from a diversity of sources and different surveys of distant galaxies.  
Most of these are deep Hubble Space
Telescope imaging surveys which have accurately  measured stellar
masses, redshifts and merger fractions out to these
redshifts.  

The galaxy merger data we use in this study come from several studies of 
the merger 
history using the CAS structural method (Conselice 2003, 2014), as well 
as galaxies in pairs (Lopez-Sanjuan et al. 2010; Bluck et al. 2009; 2012, 
Man et al. 2012). The surveys we take our merger data from include the GOODS
NICMOS Survey (Conselice et al. 2011), NICMOS imaging of 
the COSMOS field (Man et al. 2012), and the Hubble Ultra Deep
and Deep Fields (e.g., Williams et al. 1996; Conselice et al.
2008).   We also take results from  Bluck et al.
(2009, 2012) for M$_{*} > 10^{11}$ \solm galaxies from the GOODS
NICMOS Survey to $z = 3$ (Mortlock et al. 2011).  
For the most massive galaxies
with M$_{*} > 10^{11}$ \solm at $z < 3$ we supplement our data with
pair fractions taken from the Ultra Deep Survey (UDS) when fitting
with models.
 The other previous studies we use for comparison
are those from Conselice et al. (2008) and Conselice et al. (2009) who
computed the merger history for observed galaxies based on data observed
in the Hubble Ultra Deep Field, and the COSMOS and EGS fields for systems
at $z < 1.5$.    We also use new CANDELS observations of the merger
history from asymmetries calculated within the CANDELS area of
the Ultra Deep Survey field (Mortlock et al. 2013).
 
There are many other potential merger histories that we can use, but do not,
due to time-scale and major/minor merger sensitivity, we only use galaxies
in pairs and those measured with the CAS system, which has a well defined
merger time-scale for gas rich major mergers discussed in \S 2.2.  When 
examining the merger
fraction history it is clear that there are significant differences between
the various methods of measuring mergers (e.g., Conselice et al. 2009;
Lotz et al. 2011).  This is due to different methods being sensitive to
various time-scales of the merger process as well as to the 
mass ratios of mergers. 

We are interested in a specific type of merger in this paper -- systems 
which are merging and which have a progenitor mass ratio of
1:4 or lower and have merged within the past 0.4 Gyr. Detailed simulations
from Conselice (2006) \& Lotz et al. (2010a) show that a positive merger
signature is seen when the mass ratio is 1:4 in total mass.   We also 
investigate the corresponding stellar mass ratios, 
based on the dark matter
halo merger ratio, when comparing predicted halo merger fractions to observed
pair fractions (\S 5.2).
We therefore only use
the CAS mergers which have this time-scale sensitivity (e.g., Conselice
2006; Lotz et al. 2008), and pairs of galaxies where the merger mass
ratio can be measured directly and where pairs with separations of
$< 30$ kpc mergers have a similar
time-scale as the CAS selected mergers (e.g., Conselice et al. 2009; 
Bluck et al. 2012) of about $\tau_{\rm} = 0.4$ Gyr.    We also utilize 
these studies as they use similar or the same methods to calculate 
photometric redshifts and stellar masses allowing us to minimize these 
sources of uncertainty between various results.

The stellar masses we use to constrain our observed sample originate from
fitting observed SEDs to 
stellar population synthesis methods.  The procedure for this differs
between the various studies, but the results are largely consistent, and
are normalized such that they use the same IMF (Salpeter), and the
same range of stellar ranges from 0.1 to 100 \solm. 
The fitting method for our stellar masses consists of fitting a grid of 
model SEDs constructed
from Bruzual \& Charlot (2003) (BC03) stellar population synthesis models, 
using a variety of  exponentially declining star formation histories,
with various ages, metallicities and dust contents included.  The models 
we use are parameterized
by an age, and an e-folding time for parameterizing the star formation 
history, where SFR(t) $\sim\, e^{-\frac{t}{\tau}}$.  

The values of $\tau$ are
randomly selected from a range between 0.01 and 10 Gyr, while the age
of the onset of star formation ranges from 0 to 10 Gyr. The metallicity
ranges from 0.0001 to 0.05 (BC03), and the dust content is parameterized
by $\tau_{\rm v}$, the effective V-band optical depth for which we use values
$\tau_{\rm v} = 0, 0.4, 0.8, 1.0, 1.33, 1.66, 2, 2.5,5.0$.   
Although we vary several parameters,
the resulting stellar masses from our fits do not depend strongly on the
various selection criteria used to characterize the age and the metallicity
of the stellar population.

We also utilize photometric and spectroscopic redshifts for our measurements
of the stellar masses which also come into play when measuring the
evolving merger history.  The typical photometric redshift accuracy for these
surveys is quite good, with values of $\delta z/(1+z) \sim 0.03$ 
(Hartley et al. 2013; Mortlock et al. 2013).
Details of how these photometric redshifts are computed are included in the
above cited papers (e.g., Conselice et al. 2009; Mortlock et al. 2011;
Bluck et al. 2012; Mortlock et al. 2013).  The errors in the merger fractions
which we later use to fit to the predicted halo and galaxy mergers fully
take into account the uncertainties in the stellar masses and redshifts for
these samples.  This is in fact the largest source of uncertainty when
calculating the most likely cosmological model based on the merger fraction
evolution.

\section{Halo and Galaxy Merger Predictions}

In the following sections we investigate the halo and galaxy merger histories
both predicted, for halos (\S 3.1 - 3.3) and galaxies (\S 4.2), and 
in \S 4 for the observed galaxies.    In this paper we discuss many 
observed and predicted galaxy and halo mergers. To simply things we give
a brief over-view here of what is presented later.  In this section we only 
discuss halo mergers, and how they evolve as a function of halo mass, 
redshift and cosmology.   These halo mergers are the basis for the rest of the paper.
We give a detailed description of the halo merger history within 
the $\Lambda$CDM frame-work, comparing with predictions from semi-analytical 
models based on the Millennium simulations I and II, as well as with halo 
occupation distribution (HOD) modeling predictions.

In \S4 we discuss 
the observed and predicted merger history for galaxies at a given
stellar mass selection  using several methods and simulations
(e.g., Bertone \& Conselice 2009; Hopkins et al. 2010).   \S 4.1 gives
an overview of the observed galaxy merger history which we later compare
the halo mergers described in this section. \S 4.2 is a discussion of
the latest in galaxy merger history predictions from both semi-analytical
models and abundance matching.  We describe there how the predicted galaxy
mergers are unable to match data as well as the dark matter halos themselves
and therefore use a methodology to compare directly the halo mergers
to the observed galaxy mergers in \S 5.
  
\subsection{Halo Mergers as a Function of Redshift and Mass}

We first describe the merger histories of halos using the
concordance cosmology, defined as $\Omega_{\Lambda} = 0.7$,
$\Omega_{\rm m} = 0.3$ and $\sigma_{8} = 0.9$, as a function of halo 
mass.  The merger histories
for galaxy halos has been studied previously, but mostly only within
the standard cosmology.  Early results (e.g., Gottlober et al.
2001) found that the
merger history for halos can be described as a power-law increase with
redshift as $\sim (1+z)^{3}$.  More detailed predictions have been
provided by modern simulations such as the Millennium simulation 
(e.g., Bertone \& Conselice 2009; Fakhouri \& Ma 2008;
Fakhouri et al. 2010; Hopkins et al. 2010) where both the merger history 
for halos and galaxies are simulated and predicted.  These simulations find
that the merger history of halos increases as a power-law (e.g.,
Fakhouri \& Ma 2008).  These simulations also find that
galaxy mergers are less common than halo mergers, and that the
predicted galaxy merger fraction is much lower than what is 
observed (e.g., Conselice et al. 2003; Bertone \& Conselice 2009; Jogee
et al. 2009; Hopkins et al. 2010).

However, there is a better match between galaxy
mergers and halos when using
HOD models which match dark matter halos predicted to exist at
a given redshift to observed galaxies
selected by stellar mass and clustering (e.g., Hopkins et al. 2010). 
In this case the galaxy merger fraction based on stellar mass is
determined by the merger fraction of the halos in which these
galaxies are located (\S 4.2).

\subsubsection{Redshift Evolution}

Here we investigate the predicted merger history for halos using the
basic CDM model predictions with the methods outlined in \S 2.
The predicted merger histories of halos of a given halo mass limit 
(M$_{\rm halo}$) using the formalism from \S 2 are shown in Figure~1 
for a standard $\Lambda$CDM cosmology.  We select these halos 
through minimum halo mass cuts, and show results for systems 
with log M$_{\rm halo} > 9$ up to log M$_{\rm halo} > 13$.  
We only investigate here, as explained in detailed in \S 2, the 
merger fraction for these halos of a given mass which merged with another 
halo, with a mass at least a fourth of the mass of its halo, and 
within 0.4 Gyr.

\begin{figure}
 \vbox to 120mm{
\includegraphics[angle=0, width=90mm]{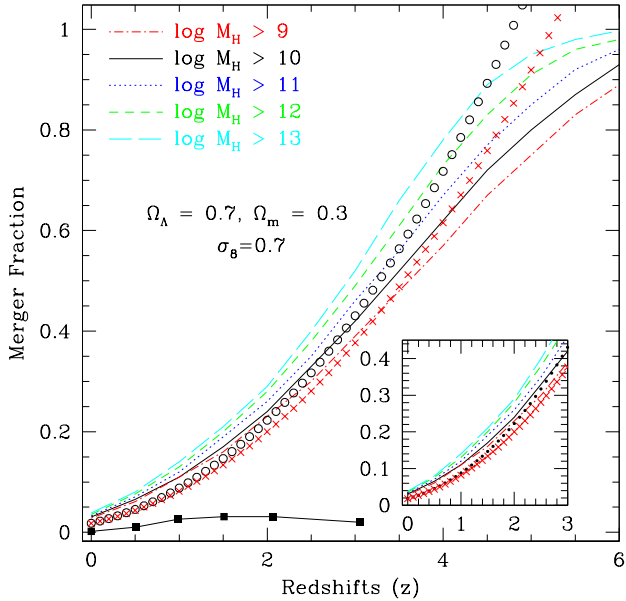}
 \caption{The halo merger history for galaxies of a given halo mass in the
concordance cosmology, with the
lines for the different halo masses defined on the upper left.  The halo merger 
fraction 
plotted here is the fraction of galaxies at a given halo mass which has merged
with another halo of at least a factor of 0.25 or greater than the mass of the progenitor halo, and within the past 0.4 Gyr.   
 The inset shows the detailed merger history for the same systems at $z < 3$.  We also show as the line
with open circles the best-fit power-law to the
merger history from the Millennium simulation (e.g., Fakhouri et al. 2010).
Furthermore, the lower solid line with box points is the Millennium simulation
prediction for galaxy mergers 
(as opposed to halo mergers) with log M$_{*} > 10$.  The red
crosses are model predictions from Stewart et al. (2009) for galaxies with
halo masses of M$_{\rm halo} = 10^{12}$ \solm.   }
} \label{sample-figure}
\vspace{3cm}
\end{figure}

Figure~1 shows a general trend such that the merger fractions of halos are 
very high in the early universe, up to redshifts $z \sim 6$, with halos of the 
highest masses at \mhalo $> 10^{13}$ \solm having a merger fraction  
close to $f_{\rm halo} \sim 1$.   Our predictions for the halo merger history
within the mass ranges where galaxies are found agree with those
predicted in the Millennium simulations I and II (Fakhouri et al. 2010), and
within the merger simulations of Stewart et al. (2009).   We
also show the predicted quantities for halo mass mergers as the 
lower line with points in Figure~1, based on the Millennium simulation, 
demonstrating the much lower values for galaxies than halos in the same 
simulation (\S 4).

We parameterize these halo merger histories as a power-law of the form (e.g., Bluck et al. 2009, 2012),

\begin{equation}
f_{\rm halo} (z, M_{\rm halo}) = f_{\rm halo}(0, M_{\rm halo}) \times (1+z)^{m},
\end{equation}

\noindent where $f_{\rm halo}(z=0, M_{\rm halo})$ is the halo merger 
fraction at $z = 0$ and $m$ is the 
power-law index for the merger history.  Higher values of $m$ are fit for
steeper merger fractions.  
Merger histories have often been fit with these power-law forms for 
galaxies at various stellar mass and luminosity cuts using real 
data for some time, although this is one of the first times this 
has been done for dark matter halos as a function of cosmology.  
Previously, Gottloeber et al. (2001) applied this fitting method to merger 
histories in the standard cosmology.


We list the results of this fitting for CDM halos of various masses in 
Table~1. These results show that
the halo merger history is similar at $z = 0$ at all
halo masses, differing by only a small amount.  However, the
power-law increase is such that the more massive galaxies have
a steeper rise in their halo mergers, and thus a larger merger
fraction at earlier times than lower mass systems.  This is
a demonstration of the hierarchical nature of structure formation
in halos.  This is also opposite to what is seen in the galaxy population
where the more massive systems appear to end their merger and formation
processes before lower mass ones (e.g., Conselice et al. 2003; Bundy
et al. 2006; Conselice et al. 2009; Mortlock et al. 2011).

\begin{table}
\centering
\begin{tabular}{c c c}
\hline \hline
Mass limit &  $f_{0}$ & $m$ \\
\hline
log M$_{\rm halo} >$ 9 & 0.030$\pm$0.001  & 1.85$\pm$0.03 \\ 
log M$_{\rm halo} >$ 10 & 0.029$\pm$0.001  & 1.94$\pm$0.03  \\ 
log M$_{\rm halo} >$ 11 &  0.033§$\pm$0.001 & 1.88$\pm$0.03 \\ 
log M$_{\rm halo} >$ 12 &  0.030$\pm$0.010 & 1.91$\pm$0.01  \\ 
log M$_{\rm halo} >$ 13 &  0.036$\pm$0.001 & 1.92$\pm$0.03  \\ 
\hline
\end{tabular}
\caption{The best fitting power-law fits from eq. (9) to the merger history for
dark matter halos using halos of different masses.  These merger
histories are listed at the halo mass M$_{\rm halo}$ limit. These fits
are for the merger history up to $z=4$.    } 
\end{table}

\subsubsection{Variation with Halo Mass}

Another remarkable aspect of Figure~1 is that the differential between
the halo mass merger histories amongst the various mass selections
is smallest at the lower redshifts, and highest
at $z > 4$. This shows that the merger properties for halos of different
halo masses is similar at lower redshifts, but diverges more at the highest
redshifts.  Therefore the halo assembly history is more distinct earlier,
as a function of M$_{\rm halo}$, rather than later, in the universe.

Currently the observed merger history is largely limited to studies at
relatively modest redshifts, those at $z < 3$ (e.g., Conselice et al. 2008, 
Mortlock et al. 2013; c.f. Conselice \& Arnold 2009).  Therefore we, in 
particular, examine within this paper the
halo merger histories given our cosmological models at these lower
redshifts, limited to $z < 3$.  At these redshifts the merger fractions for
halos over four orders of magnitude in mass differ 
by a maximum of $\delta f_{\rm halo} \sim 0.2$.

To study this in more detail in the inset of Figure~1 we show 
merger histories for halos as a function of redshift
up to $z \sim 3$.  To quantify this, we fit the change in the halo 
merger history as a function of halo mass at two redshifts,
$z = 2.5$ and $z = 1$.  We later use these to determine the 
uncertainty in the matching of halo and stellar
masses, and therefore to go from the halo merger history to the galaxy
merger history.  At all redshifts we find that this relation is linear and
is well described by a function of the form:

\begin{equation}
f_{\rm halo} (z, M_{\rm halo}) = a \times {\rm log M_{*}} + b.
\end{equation}

\noindent We find for $z=2.5$ the values $a = 0.025\pm0.001$ and $b
= 0.08\pm0.01$.  This is a very shallow slope, and shows that the halo merger 
fraction at a given redshift does not change very much between halo masses
of log M$_{\rm halo}$ = 9-13.    At lower redshifts, the relation becomes
even flatter, with fits for $z = 1$, $a = 0.010\pm0.001$ and $b =
0.011\pm0.002$.  For the remainder of the paper we only examine the
merger history out to $z = 3$ as this is currently where we have the most 
certainty in our ability to measure masses and the merger history in 
actual galaxies (\S 4).

\subsection{Variation with $\Omega_{\rm m}$ and $\Omega_{\Lambda}$}

One of the features we investigate with our halo merger histories is
how the merger history changes within different underlying cosmological 
parameters.  This allows us to investigate, among other things, whether the merger 
history of galaxies is consistent with the dominant cosmological model.
In Figure~2 we show the merger history for halos of masses
\mhalo $> 10^{12}$ \solm and for \mhalo $> 10^{11}$~\solm using our 
prescribed method for finding and defining merging halos (\S 2), and
within the different cosmologies listed in Table~2.

What Figure~2 shows is that the cosmology built into the structural evolution
of the universe can potentially strongly alter the history of halo mergers and thus the formation of structure.  
The lowest merger histories are those with a cosmology which has a 
zero cosmological
constant, and a low dark matter content.  The highest merger histories are
for Einstein-de Sitter cosmologies in which the the total matter 
density is $\Omega_{\rm m} = 1$ with a zero cosmological constant (Cosmo4 in
Table~2).    Furthermore, there is a clear correlation with higher merger fractions
for higher values of $\Omega_{\rm tot}$.  This is an indication that the
merger history is tracing to some degree the geometry of the universe.  

This figure however shows that even for a total density of $\Omega
=1$ there are variations within the merger history.  Three extreme models
are shown in Figure~2 -- one in which the energy density is completely in the
form of matter (Cosmo-4; the short dashed line), one which has the currently
accepted cosmological model (Cosmo-7; \om = 0.3, \ol = 0.7; $\sigma_{8} = 0.9$ the 
dot-dashed line), and one in which
the energy density is divided evenly between $\Omega_{\rm m}$ and 
$\Omega_{\Lambda}$ (Cosmo5; long
dashed line).  

In general, the higher the matter density, the higher the merger fraction 
within the halo merger models.  This is parameterized in \S 5.2 in terms of 
the value of $\Omega_{\Lambda}$.  The halo merger fraction begins 
to turn over when there is a
higher $\Lambda$ term in the cosmology.   This is as expected, given that 
detailed numerical models of galaxy formation have shown for many years
that the large scale-structure of the universe depends strongly upon the
assumed cosmology, as well as the temperature of the dark matter particle
(e.g., Jenkins et al. 1998; Menci et al. 2012).  

We later examine in \S 5 how the halo merger history relates to the galaxy
merger history, and how comparisons between the two can potentially reveal
information concerning cosmological parameters, as well as information 
regarding the measured galaxy merger history.  

We parameterize the halo merger evolution for different cosmologies using
the same power-law formalism explained in \S 3.1.  These resulting fits
are listed in Table~2 along with their name and associated
cosmological parameters.  Just as for galaxies of different halo masses, we
find that the values of $f_{0}$ do not vary significantly between the 
different cosmologies, and that the most variation is within the 
values of $m$.  In general we find that the higher the value 
of $\Omega_{\Lambda}$, the higher the fitted value of the 
power-law slope, $m$.  The average $m$ value for $\Omega_{\Lambda} = 0.7$
is $m = 1.93$, while for $\Omega_{\Lambda} = 0.3$ it is $m = 1.76$.  The 
decline
in mergers is therefore steeper for higher values of $\Lambda$ due to the
accelerated expansion, lowering the number of mergers at a faster rate.

\begin{table}
\centering
\begin{tabular}{c c c c c c c}
\hline \hline
Model & $\Omega_{\Lambda}$ & $\Omega_{\rm m}$ & $\Omega_{\rm tot}$ & $\sigma_{8}$ & $f_{0}$ & $m$ \\
\hline
Cosmo-1 & 0.0 & 0.1 & 0.1 & 0.9 & 0.027  & 1.73 \\ 
Cosmo-2 & 0.7 & 0.3 & 1.0 & 0.9 & 0.032  & 1.92 \\ 
Cosmo-3 & 0.0 & 0.3 & 0.3 & 0.9 & 0.035 & 1.82 \\ 
Cosmo-4 & 0.0 & 1.0 & 1.0 & 0.9 & 0.052 & 1.73 \\ 
Cosmo-5 & 0.5 & 0.5 & 1.0 & 0.9 & 0.041 & 1.88  \\ 
Cosmo-6 & 0.7 & 0.3 & 1.0 & 0.7 & 0.038 & 1.93  \\ 
Cosmo-7 & 0.7 & 0.3 & 1.0 & 0.8 & 0.034 & 1.97  \\ 
Cosmo-8 & 0.7 & 0.3 & 1.0 & 1.0 & 0.032 & 1.94   \\ 
Cosmo-9 & 0.7 & 0.3 & 1.0 & 1.1 & 0.033 & 1.88 \\ 
\hline
\end{tabular}
\caption{The cosmological models we use in this paper to compare with the
observed merger fractions and their various fitted parameters.   These are for
galaxies with halo masses of M$_{\rm halo} > 10^{12}$ \solm. Listed are each model's $\Omega_{\Lambda}$,
$\Omega_{\rm m}$ and $\sigma_{\rm 8}$ values.  We also show the best
fitting $f_{0}$ and $m$ values for the power-law fits as discussed in
\S 3.1.      } 
\end{table}

\begin{table}
\centering
\begin{tabular}{r c c c}
\hline \hline
Mass Ranges & Dark Matter & $f_{0}$ & $m$ \\
\hline
$9 < {\rm log} {\rm M_{*}} < 10$ & CDM & 0.007$\pm$0.002 & 1.77$\pm$0.43 \\
 log ${\rm M_{*}} > 10$ &  CDM &  0.005$\pm$0.002 & 2.00$\pm$0.56 \\
 log ${\rm M_{*}} > 11$ &  CDM & 0.030$\pm$0.004 & 1.19$\pm$0.21 \\
 log ${\rm M_{*}} > 11.5$ &  CDM & 0.050$\pm$0.010 & 0.99$\pm$0.14 \\
$9 <  {\rm log} {\rm M_{*}} < 10$ & WDM & 0.010$\pm$0.002 & 1.40$\pm$0.21 \\
 log ${\rm M_{*}} > 10$ &  WDM & 0.006$\pm$0.001 & 2.07$\pm$0.13 \\
 log ${\rm M_{*}} > 10.5$ &  WDM & 0.017$\pm$0.003 & 2.04$\pm$0.14 \\
 log ${\rm M_{*}} > 11$ & WDM &  0.040$\pm$0.010 & 1.89$\pm$0.15 \\
\hline
\end{tabular}
\caption{The best fit power-law parameters in the form 
$f_{\rm m} = f_{0} \times (1+z)^{m}$ based on the output from the galaxy 
merger simulations based on the Millennium
Simulation and the WDM simulations of Menci et al. (2012).   } 
\end{table}

\begin{table}
\centering
\begin{tabular}{r c c}
\hline \hline
$\omega$ & $f_{0}$ & $m$ \\
\hline
$-0.33$ & 0.0380$\pm$0.0001 & 1.634$\pm$0.003 \\
$-0.40$ & 0.0374$\pm$0.0002 & 1.661$\pm$0.004 \\
$-0.50$ & 0.0360$\pm$0.0003 & 1.705$\pm$0.007 \\
$-0.60$ & 0.0344$\pm$0.0003 & 1.749$\pm$0.007 \\
$-0.70$ & 0.0331$\pm$0.0003 & 1.791$\pm$0.007 \\
$-0.80$ & 0.0320$\pm$0.0002 & 1.826$\pm$0.005 \\
$-0.90$ & 0.0312$\pm$0.0001 & 1.855$\pm$0.004 \\
$-1.00$ & 0.0307$\pm$0.0001 & 1.878$\pm$0.002 \\
$-1.10$ & 0.0304$\pm$0.0001 & 1.895$\pm$0.002 \\
$-1.20$ & 0.0302$\pm$0.0001 & 1.908$\pm$0.004 \\
\hline
\end{tabular}
\caption{The best fit power-law parameters in the form 
$f_{\rm m} = f_{0} \times (1+z)^{m}$ based on fits to dark halo
mergers with different values of $\omega$.  These fits are for
halos with M$_{\rm halo} > 10^{11}$. } 
\end{table}

\begin{figure*}
 \vbox to 120mm{
\includegraphics[angle=0, width=180mm]{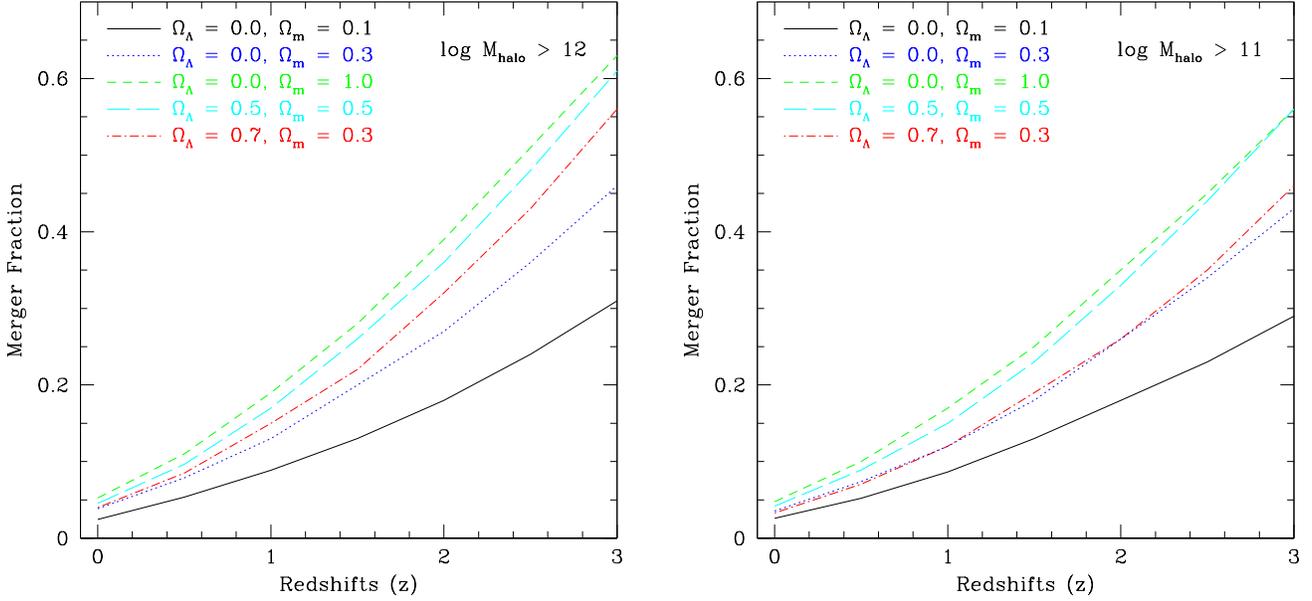}
 \caption{The halo merger histories for a) halos with masses \mhalo $> 10^{12}$ \solm and b) halos with masses \mhalo $> 10^{11}$ \solm.  The various lines show how the merger history varies with redshifts for halos of these given masses and using the cosmologies shown in the upper left.  In general we find that cosmologies with the highest matter densities have
the highest merger fractions, although higher values of $\Lambda$ produce
a smaller merger fraction at a given redshift.  }
} \label{sample-figure}
\end{figure*}

\subsection{Halo Merger Dependence on $\sigma_{8}$}

One cosmological parameter that can vary, and depends on
large scale structure, and therefore also the halo and galaxy 
formation history, is the value of $\sigma_{8}$, the normalization 
of the matter power spectrum, as measured in the RMS dispersion 
of total mass density within 8 Mpc spheres.  

We show the variation of the halo merger histories using the standard
cosmology of \om = 0.3, \ol = 0.7 but with different variations of the
value of $\sigma_{8}$ in Figure~3.   The differences in the predicted
halo merger history between these
values of $\sigma_{8}$ are smallest at $z < 1$, where the merger
fraction only varies by $\delta f_{\rm halo} \sim 0.02$ over 
the range of $\sigma_{8}$
= 0.7 to 1.1.  For the $\sigma_{8}$ = 0.7 models we find at $z = 1$
that the halo merger fraction is $f_{\rm halo} = 0.15$, while for 
$\sigma_{8}$ = 1.1 the halo merger fraction is $f_{\rm halo} = 0.12$.  
This implies that at a given redshift the accuracy of our merger 
fractions would have to be better than a few percentage, which is 
easier to accomplish than that required to distinguish between 
various values of likely different $\Omega$ cosmologies (\S 5).

We tabulate in Table~2 the values of the best fit power-laws to these
merger histories at various values of $\sigma_{8}$. 
We find a strong linear relationship between the halo merger fraction
and the value of $\sigma_{8}$, such that,

\begin{equation}
f_{\rm halo} = (-0.230\pm0.06)\times \sigma_{8} + (0.592\pm0.006),
\end{equation}

\noindent at a redshift of $z = 2.5$.  This demonstrates that there
are higher values of both $f_{0}$ and $m$ for universes
with higher values of $\sigma_{8}$.  This implies that when the RMS
fluctuations of spheres of dark matter with a radius of 8  h$^{-1}$ Mpc are
smaller, there is a higher rate of merging within the universe.   This
result is likely due to the fact that the value of $\sigma_{8}$ is directly
tied to the halo mass function.  That is, when $\sigma_{8}$ increases this
effectively shifts the halo mass function to higher values.  This then
increases the normalization needed to reach the same vale of the merger
rate as at a lower value of $\sigma_{8}$.  From Table~1 we can see
that lower mass halos have a lower merger rate and thus a higher $\sigma_{8}$
pushes the effective scaling lower, thus creating a lower merger rate
at a given halo mass.

\begin{table}
\centering
\begin{tabular}{c c c}
\hline \hline
$\sigma_{8}$ &  $f_{0}$ & $m$ \\
\hline
0.7 & 0.038$\pm$0.001  & 1.93$\pm$0.02 \\ 
0.8 & 0.036$\pm$0.001  & 1.92$\pm$0.02  \\ 
1.0 &  0.033§$\pm$0.001 & 1.91$\pm$0.01 \\ 
1.1 &  0.032$\pm$0.001 & 1.88$\pm$0.01  \\ 
\hline
\end{tabular}
\caption{The best fitting power-law fits to the merger history for
dark matter halos at masses M$_{\rm halo} > 10^{12}$ \solm at $z = 2.5$
as a function of $\sigma_{8}$.     } 
\end{table}

\begin{figure}
 \vbox to 120mm{
\includegraphics[angle=0, width=90mm]{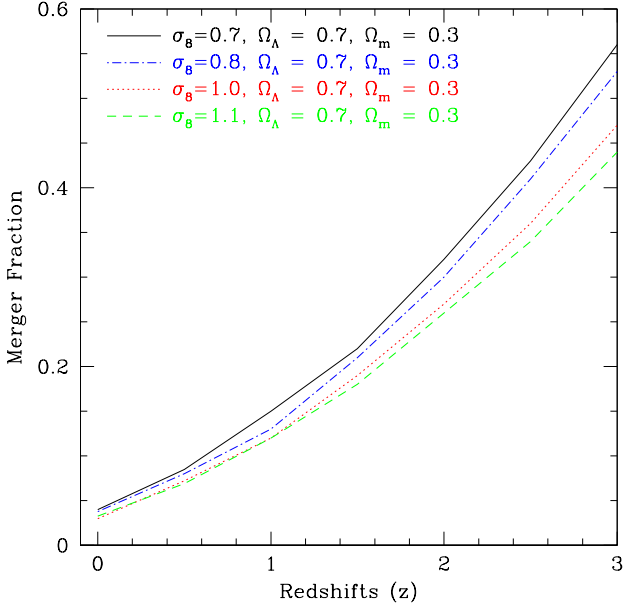}
 \caption{The evolution of the halo merger fraction for halos with
\mhalo $> 10^{12}$ \solm as a function of the value of $\sigma_{8}$.
The range we show here is for $\sigma_{8}$ = 0.7 to $\sigma_{8} = 1.1$,
with the highest halo merger fractions for those evolving within
a universe that has the lowest $\sigma_{8}$.  }
} \label{sample-figure}
\end{figure} 
\vspace{10cm}

\section{Comparison of Observed Galaxy Mergers to Simulated Galaxy Mergers}

In this section we discuss the comparison of our observations of the
galaxy merger history with the theoretically simulated galaxy
merger histories.  We do this before we examine the comparison of halo
merger histories to galaxies as comparing with simulated galaxy mergers
is a more direct comparison, although it largely fails as we show.

\subsection{Background}

While the galaxy merger history as predicted in various simulations has
been discussed in detail elsewhere (e.g., Jogee et al. 2009;
Bertone \& Conselice 2009; Hopkins et al. 2010), we
give a short summary here, as well as a comparison to the halo models.
We show on Figure~1 as the solid dark line towards the lower part of the
plot the galaxy
merger history for systems selected by a stellar mass limit of log M$_{*} > 10$
as predicted in the Millennium simulation.  This roughly corresponds to
a halo mass a factor of ten higher, based on the results from 
Twite et al. (2014) who investigate the ratios of stellar and
halo masses for galaxies (see \S 5.1).  In general, it appears that
the predicted galaxy merger history is much lower than the
halo merger history at roughly the same selection.  The ratio between
the halo and galaxy merger fractions from the Millennium simulation 
is roughly constant, at a ratio of
$\sim 7.5$ up to  redshifts of $z \sim 3$.

We directly compare the predicted galaxy merger history from the
Millennium simulation to merger fraction data in Figure~4 \& 5.  
The first piece of evidence
to take away from this figure is that the predicted galaxy merger histories are
quite different from the observed galaxy merger histories.  In fact, as 
discussed in e.g., Jogee et al. (2009) and Bertone \& Conselice (2009) the
merger histories for galaxies does not match the observations, with too
few major merger observed than what is predicted in the galaxy merger
histories from the Millennium simulation (Figure~4\&5), although
in \S 4.1.2 we discuss how this changes when using Warm Dark Matter galaxy
merger simulations.

An important issue is that the galaxy merger history can, and does,
depend strongly on the galaxy formation model used.  We show this in
detail in Figure~4 where four different realizations of the Millennium
simulation are shown from Bertone et al. (2007), Bower et al. (2006),
De Lucia et al. (2006) and Guo et al. (2010).  As can be seen, these
models do not agree with the observations.   It is interesting as well that 
the predicted galaxy mergers within different realizations of
the Millennium simulation tend to agree with each other for 
M$_{*} > 10^{10}$ \solm galaxies, but widely disagree for
those with masses in the M$_{*} > 10^{11}$ \solm range.  The reason for this is due to
these models not agreeing on which galaxies belong in which halo, which
becomes more of an important issue at higher halo masses.  These
massive galaxies are on the exponential tail of the mass function and
therefore any differences in the star formation history or feedback
can produce significant differences in the merger history through
matching which galaxy is in which halo (see also Bertone \&
Conselice 2009).

\begin{figure*}
 \vbox to 140mm{
\includegraphics[angle=0, width=180mm]{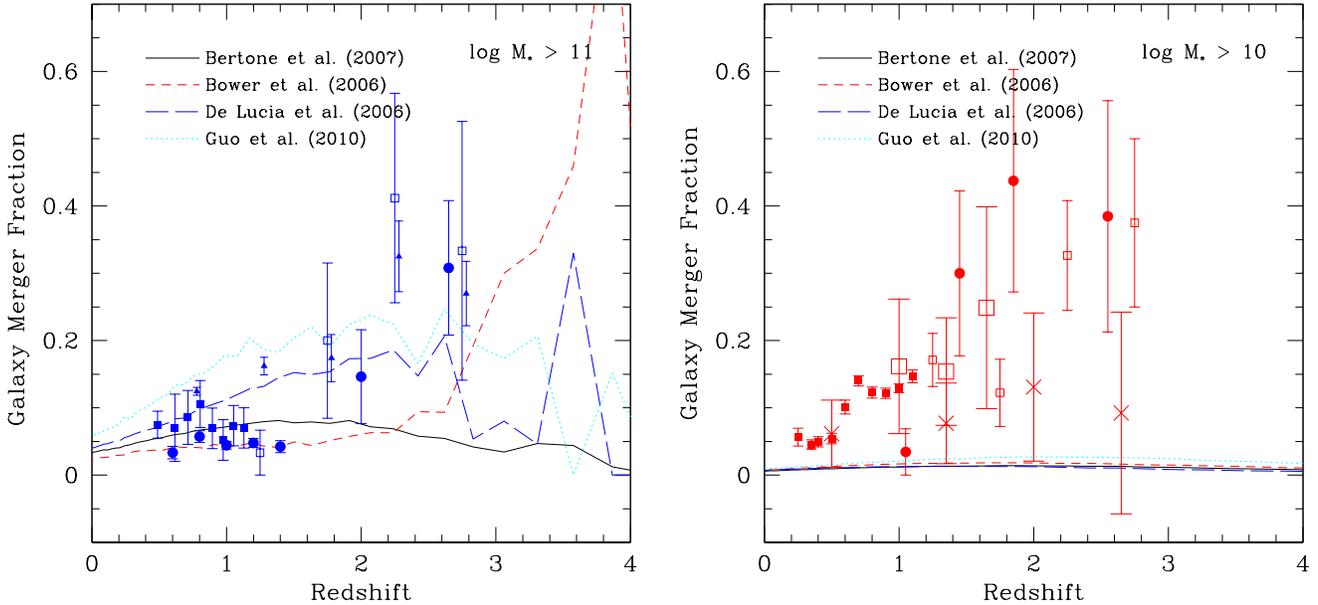}
 \caption{The predicted merger history for galaxies using different
realizations of the semi-analytical Millennium I simulation.  Shown
here are models from Bertone et al. (2007) (solid line), Bower
et al. (2006) (dashed red line), De Lucia et al. (2006) (long-dashed
blue line), and Guo et al. (2010) (cyan dotted line). The large
variation in models seen for the M$_{*} > 10^{11}$ \solm galaxies
is solely due to the identification of galaxies with halos, as the
underlying merger history for the halos in these simulations
are identical.   The points with error bars shown are
actual measures of the merger fraction for galaxies as a function of
redshift.  The blue points on the left show the merger history 
for M$_{*} > 10^{11}$ \solm systems including results from Conselice
et al. (2009) (solid boxes at $z < 1.2$; Bluck et al. (2009, 2012)
(solid circles at $z > 0.5$); and Mortlock et al. (2013) (open boxes
at $z > 1$).  The red points on the right are for  M$_{*} > 10^{10}$ \solm
galaxies, including results from Conselice et al. (2003) at $z > 1$ in
the HDF (solid circles); Conselice et al. (2009) (solid boxes at
$z < 1.2$); and Mortlock et al. (2013) (small open boxes at $z > 1$).   Also 
shown are pair merger fractions at separations of $< 30$ kpc: (Man et al.
2012; crosses at $z > 1$); Lopez-Sanjuan et al. (2010) (large open boxes). 
These points are used in later figures as well.
 }
\vspace{3cm}
} \label{sample-figure}
\end{figure*}

\subsubsection{Detailed Comparison of Models with Data}

With one exception, the observed galaxy merger histories with redshift 
start at low values, peak at redshifts of around $z \sim 2$, 
and then decline thereafter at higher redshifts. This is similar 
to what is found within the Millennium simulation predictions for galaxies,
although often at a lower level.  In contrast,
the merger history for halos continues to steeply increase at higher
redshifts at all masses (Figure~1).  

This difference is very likely due to the way
sub-halos are dealt with within the Millennium simulation.  When
a galaxy is accreted into a larger halo it losses all of its cold gas,
and therefore cannot produce new stars and thus when the merger occurs
after some dynamical friction time-scale, the mass ratio of the merger
is low.  These mergers thereby end up as minor mergers, although in some
ways in baryons these would still be major mergers if the striped gas
was included.  In fact, the high minor-merger fraction predicted in the
Millennium simulation suggests that this might be occurring (Bertone \&
Conselice 2009). 
Furthermore, simple computational differences, such as using different
methods to calculate the feedback from supernova, can dramatically change the
measured galaxy merger history (Bertone \& Conselice 2009).  Investigating
in more detail these differences is important, but we do not discuss
this issue further in this paper.

\begin{figure*}
 \vbox to 120mm{
\includegraphics[angle=0, width=180mm]{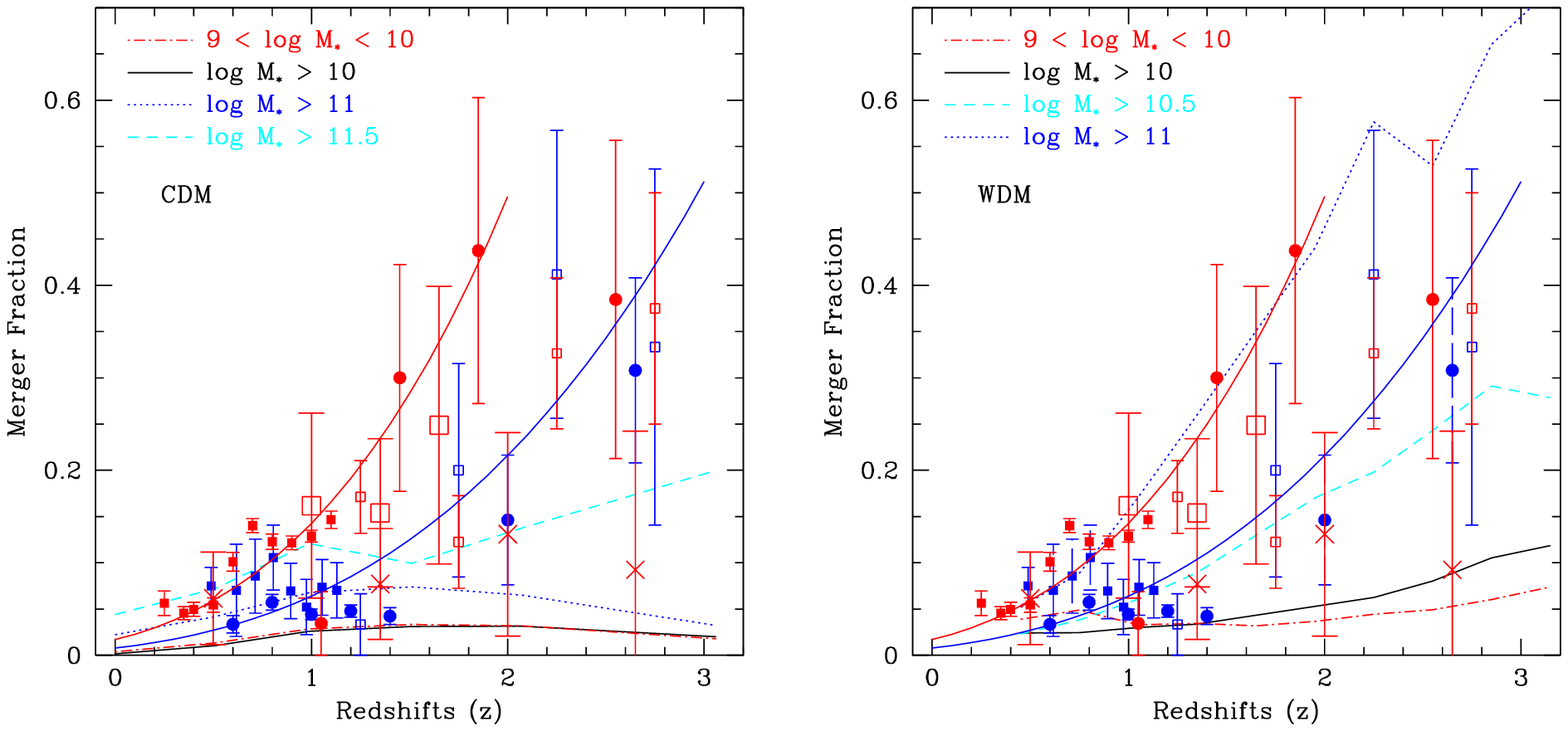}
 \caption{The merger history for  galaxies using two simulations utilizing
different dark matter particle temperatures out to $z = 3$.  The left panel 
shows the Semi-analytical merger history predictions for galaxies with various 
stellar masses as taken from the Millennium simulation
results of Bertone et al. (2007).  The various lines show the
merger fraction, as measured using the same criteria as used to select
 the merger
histories for the halos shown in in Figure~2.  The right panel shows
the merger history at the same stellar mass limits but with using
a simulation with Warm Dark Matter (Menci et al. 2012).  The solid red line
is the best fit power-law to the observed evolution in galaxy mergers for
the M$_{*} > 10^{10}$ \solm galaxies, while the solid blue line shows the 
corresponding best fit for the M$_{*} > 10^{11}$ \solm galaxies.  The
data points used for comparison are the same as in Figure~4.}
} \label{sample-figure}
\end{figure*}

To compare these different observed merger histories, for halos and galaxies
and the actual merger histories, in a more quantitative way, we characterize 
the merger histories by fitting the predicted galaxy merger histories 
with the power-law form given by equation (9).  The results of 
these fits are shown in Table~3.  Fits such as these
typically only provide a good fit to the data up to some redshift where the
merger fraction begins to turn over to lower values. However, with some 
exceptions, these are good fits to the data at $z < 3$ for both halo 
models, and the actual observed galaxy mergers.

When we fit these merger fractions to the
halo merger histories we find that their power-law indices are all
$m \sim 2$.  This differs from what we find
when we compute the same power-law indices for observed galaxies, 
with numbers ranging from $m = 3$ to $m = 1$ 
(Tables~1-3; e.g., Conselice et al. 2009; Lotz et al. 2011).  Galaxy mergers
selected in the manner we use for the halo mergers generally always find
that $m \sim 3$ (e.g., Conselice et al. 2008; Bluck et al. 2012).
We also list in Table~3 the fitted power-law parameters for both CDM and
WDM simulations which we discuss in \S 4.1.2.

Overall, we find that the halo merger history is significantly 
different from what is predicted
in galaxy merger evolution from the Millennium simulation.  In fact, the
halo merger history is a better description of the observed
galaxy merger history than the predicted galaxy mergers using the
same simulations (\S 5).     This however
shows that the fundamental merger histories for galaxies and halos can be quite
different, and new approaches of establishing the connection between halo and
galaxy mergers is needed.

While there is not a good match between semi-analytical model predictions
of galaxy mergers and the observations, when deriving the merger history
through abundance matching (e.g., Hopkins et al. 2010), there is a better
match with data compared with theory results and simulations at some halo
mass ranges (see also Figure~4 \& 5). We show this comparison for abundance
matched mergers in Figure~6 where the galaxy merger data is plotted
along side corresponding mergers derived from abundance matched samples
from Hopkins et al. (2010) as measured using
the WMAP1, WMAP3, and WMAP5 cosmological parameters.   These cosmological 
parameters are: ($\Omega_{\rm m}$, $\Omega_{\Lambda}$, 
h, $\sigma_{8}$, n$_{\rm s}$) = (0.3, 0.7, 0.7, 0.9, 1.0) for the 
concordance cosmology, ($\Omega_{\rm m}$, $\Omega_{\Lambda}$, h, 
$\sigma_{8}$, n$_{\rm s}$) = (0.27, 0.73, 0.71, 0.84, 0.96) for WMAP1, 
($\Omega_{\rm m}$, $\Omega_{\Lambda}$, h, $\sigma_{8}$, and n$_{\rm s}$) = (0.268, 0.732, 0.704, 0.84, 0.96) 
for WMAP3, ($\Omega_{\rm m}$, $\Omega_{\Lambda}$, h, $\sigma_{8}$, n$_{\rm s}$) = (0.274, 0.726, 0.705, 0.776, 0.95) 
for WMAP5.

The different lines in Figure~6 show the different cosmologies 
used to measure the merger history.  This is slightly different from 
our approach, which is to see how the merger history varies as a 
function of cosmology, as opposed to determining how the 
merger history changes due to different measured cosmological parameters. 

\begin{figure}
 \vbox to 120mm{
\includegraphics[angle=0, width=90mm]{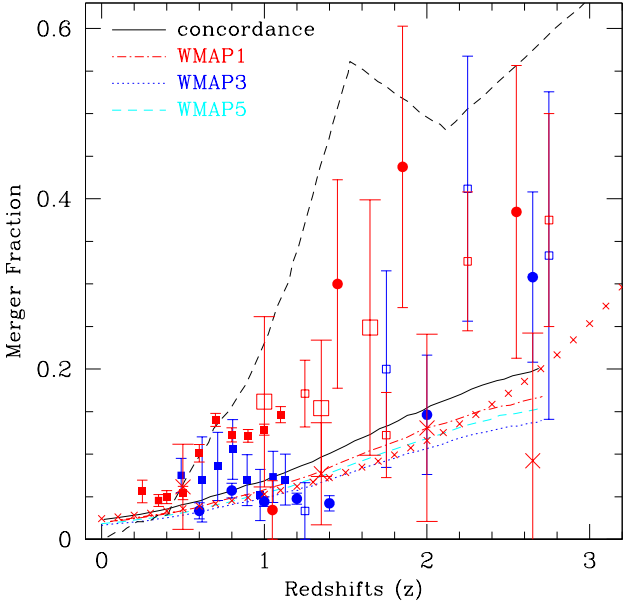}
 \caption{Plot of the observed and predicted galaxy merger history.  The
predicted lines are from the WMAP1, WMAP3 AND WMAP5 cosmologies used to
calculated the merger history in Hopkins et al. (2010).  The data with
error-bars are
the same as used in Figure~4.  The dashed line at the upper end of the
figure is from numerical models by Maller et al. (2006). }
} \label{sample-figure}
\end{figure}

As can be seen, even within these models there is a slight 
difference in the calculated merger history from abundance matching.  
The merger histories here are still lower than
what we find observationally, however the highest merger fractions are 
for the cosmologies with the highest value of $\Omega_{\rm m}$.  As 
the recent Planck cosmology has an even higher $\Omega_{\rm m}$ than 
the concordance model (Planck collaboration, Ade et al. 2013),  the 
merger history using this cosmology would be higher than those 
shown here, and would better match the data. 

The important take away message here is that the halo merger history 
does a much better job of matching the observed galaxy merger history 
than any galaxy merger history prediction.  This is a fundamental 
insight and one in which we now
expand on in \S 5, with the use of the halo merger models as
a measure of cosmology.  The
idea we explore in this paper is to not use the predicted 
galaxy formation merger histories, but the
halo merger histories and to do `inverse modeling' whereby the galaxy
sample's halo mass is derived, and then compared with halo mass merger
histories. This side-steps the need for understanding the detailed physics
in simulations of galaxy formation, but does require knowledge of how to
convert from observed galaxy to inferred halo mass.  This is essentially what
is done when using galaxy cluster as observational probes -- the 
cluster's dark matter is measured and compared with theory, especially
for cosmological tools such as galaxy cluster number density (e.g.,
Vikhlinin et al. 2009).

\subsubsection{Variation with Dark Matter Particle Temperature}

We also investigate how the galaxy merger history varies as a function of
the temperature of the dark matter particle.  This is a different topic in 
some ways from the variation in structure formation depending on cosmological 
parameters, yet this is another cosmological feature that does influence 
structure, and thus we
include it here.  Since the structure formation models we examined before all
depend on Cold Dark Matter, we utilize the Menci et al. (2012) dark matter
semi-analytical models to make the comparison with the CDM models from
the Millennium simulation.  In Figure~5 we show the
merger history for the Cold Dark Matter based Millennium simulation as
well as a Warm Dark Matter simulation from Menci et al. (2012).  

Figure~5 shows that the Warm Dark Matter simulations do a better job of 
matching the observations, which is also seen in comparisons
to observed galaxy number densities in 
the CDM and WDM simulations (e.g., Menci et al. 2012).   Effectively the 
merger fraction in WDM galaxy simulations is a factor of 1.3 higher for
systems with M$_{*} > 10^{10}$ \solm, and a factor of three higher at masses
of M$_{*} > 10^{11.5}$ \solm.  It is clearly the more massive systems 
(Figure~5)
which show the greatest difference between the CDM and WDM predictions.  As CDM
halo mergers, as well as those predicted in the Millennium 
simulation itself,
predict a higher halo merger rate similar to the observed galaxy rate, then 
it is
likely that the problem with the CDM galaxy mergers is not due to CDM itself, 
but due to how baryons are handled in these simulations.

\section{Observations of Mergers as a Cosmological Tool}

The previous section showed that observed galaxy mergers have a different
evolution than that seen in predictions of galaxy mergers.  In this
section we discuss in detail how to compare simulated halo mergers with
observed galaxy mergers.  
This is a non-trivial comparison to make, and depends on several assumptions
that we discuss in detail.  One of these issues is the
correspondence between
the halo and stellar masses which is required to match halo histories
to that of galaxies, or on just a practical level matching a halo
mass to a stellar mass.  The relationship between these two masses is
ideally well defined and has a small scatter to  minimize mismatches
between the halos and galaxies.
Other issues in a comparison to halos and galaxies have to do with 
relating the  time-scales for halo merging to
those of galaxy mergers.  

We previously discussed briefly how there is a better agreement between 
predicted halo mergers and the observed
galaxy merger fractions. One of our major conclusions is that it is better
to compare the observed galaxy merger history with these predicted halo mergers
rather than with the predicted galaxy mergers. In this section we present a new method for
comparing the observed merger history to the predicted halo merger history
as a measurement of cosmology.  We show that this can be done through several
steps, each containing a certain amount of uncertainty.  This process is
such that we convert the observed stellar mass selects sample into a halo
mass using relations that we discuss based on halo abundance matching and
kinematic data (\S 5.1).  These relations are discussed in more detail in Twite
et al. (2014).  There are uncertainties in these conversions which we
discuss and include in our analysis.   the first part of this process
is the conversion or matching of halo and stellar masses to effectively
use the halo mass mergers we discussed in \S 2-3.   We conclude this 
section with how to best compare the modeled dark matter halo mergers 
to the observed galaxy mergers, especially
in the future when better data and simulations are available.  

\subsection{Relation between Halo and Stellar Mass}

A major issue that needs to be addressed in any paper that compares
halo properties to galaxies is how to relate
the stellar mass of a galaxy to its underlying halo mass.   This
relates back to our idea of comparing with CDM, and other dark matter
based models. What we therefore need is an accurate way to relate the observed
stellar mass to the inferred halo/total masses of galaxies.

This can be done in a number of ways, including kinematics and gravitational 
lensing to derive the total masses of galaxies.  This problem has been
investigated at high redshift in several studies, including Conselice et al. 
(2005), Foucauld et al. (2010) and Twite et al. (2014).  All of these studies 
conclude that the dark matter to stellar matter ratio evolves together such 
that the different components of the assembly history are increasing at a 
similar rate.
This is an important aspect, as it allows us to compare the observed merger
history of galaxies which is based on a stellar mass selection, to that
of halo selection, which is predicted in models.  Yet another approach towards
understanding the relationship between stellar and halo masses is halo
abundance matching (e.g., Conroy et al. 2007; Twite et al. 2014) 
which we also investigate for relating stellar and halo masses.  

We calculate the abundance matching derived relation between stellar and halo
masses using the stellar mass functions from Mortlock et al. (2011).    
We match number densities from galaxies with measured stellar masses
to dark matter halo abundances at the same 
redshifts.  This allows us to associate each stellar mass range with a halo 
mass range.  This is described in more detail in Twite et al. (2014).
        
In summary, to compute this relation the mass function of dark matter 
halos (including sub-halos) is assumed to be monotonically related to 
the observed stellar mass function of galaxies with zero scatter.  This 
relation is given by,

\begin{equation}
n_g(>M_{\rm star})=n_h(>M_{\rm halo})
\end{equation}

\noindent where, the values $n_g$ and $n_h$ are the number density of 
galaxies and dark matter halos, respectively. 

We derive these values for the halos from the Jenkins et al. (2001) 
modification to the Sheth \& Tormen (1999) halo mass function using the 
analytic halo model of 
Seljak (2000).  We also generate the linear power spectrum using the 
fitting formulae of Eisenstein \& Hu (1998), the same we use to predict
the halo mergers.  The 
predicted number density of dark matter halos is then given by,

\begin{equation}
n_h(>M_{\rm halo})=\bar{\rho}\int_{M_{min}}^{\inf}\frac{<N>}{M_{\rm halo}} f(\nu)\, d\nu.
\end{equation}

\noindent Where $f(\nu)$ is the scale independent halo mass function, 
$\nu=[\delta_c/\sigma(M_{\rm{halo}})]^2$ ($\delta_c=1.68$ is the value for 
spherical over-density collapse). $\sigma(M_{\rm{halo}})$ is the variance 
in spheres of matter in the linear power spectrum, $\bar{\rho}$ is the 
mean density of the Universe, and $\left<N\right>$ is the average number 
of halos, including sub-halos where we assume the fraction of sub-halos 
(f$_{\rm{sub}}$) is described by,

\begin{equation}
f_{sub}=0.2-\frac{0.1}{3}z,
\end{equation}

\noindent as in Conroy \& Wechsler (2009). 
This method of halo abundance matching does a good job at matching 
observations at multiple epochs (e.g., mass-to-light ratios, 
clustering measurements).

The halo abundance matching  we use for our main Lambda CDM cosmology 
include errors that incorporate the difference in the abundance matching 
at the redshift bounds of each bin, and the uncertainty due errors in the 
stellar mass functions. We investigate the same abundances using the
mass function and galaxy bias as that of Tinker et al.(2008) and 
Tinker et al. (2010). These results are very close to that of the 
Jenkins et al.
(2001) halo mass function, and its resulting bias.  We also find that the 
halo to stellar mass relationship from abundance matching, using 
our main Lambda CDM cosmology,
and the cosmology used in Conroy \& Wechsler (2009) extracted using DEXTER are
almost identical at higher redshifts.
Our results also match the redshift $z=1$ Conroy \& Wechsler (2009) work 
well considering our redshift bin is slightly higher, and we use 
different galaxy stellar mass functions.
Similarly to Conroy \& Wechsler (2009) who do not trust their $z>1$ 
results, we do not place
much emphasis on our high-$z$ results below the relevant stellar mass 
limits. e.g., for $z=2$ 
at log M$_{*} > 10^{10.5}$ \solm.  The evolution in shape at the high 
mass end is also in agreement with Conroy \& Wechsler (2009). Therefore 
the range of merger histories for different cosmologies in the 
abundance matching will not vary by much.

Although this paper is not focused on abundance matching, which will be 
described in more detail in Twite et al. (2014), another issue that we 
investigate is how well our abundance matching can reproduce the 
angular correlation function of galaxies to check for halo bias within our
halo abundance matching models. It should be noted that this is intended as a 
rough check and not a robust fit of the galaxy HOD as this is not the 
focus of this work.  We produce angular correlation functions
using our HOD code with a simplified version of the 5-parameter HOD 
model (e.g., Zheng et al. 2005), where N$_{\rm centrals} = 1$ if 
M$_{\rm h} >$ M$_{\rm h}$ (M$_{*}^{\rm min}$) and M$_{\rm h}<$ 
M$_{\rm h}$ (M$_{\rm *}^{\rm max}$), and N$_{\rm sat} 
= \left[(M - M_0) / M_{1'}\right]^{\alpha}$. Here we 
have set $\alpha = 1$, $\log_{10}(M_0) =  \log_{10}(M) - 1.5$
 and $\log_{10}(M_{1'}) =  \log_{10}(M) + 1.0$, which correspond to 
average HOD fit parameters (e.g., Zehavi et al. 2011).  

We also compare the resulting correlation function with the 
power-law fits from Foucaud et al (2010). Here we have
assumed a Gaussian redshift distribution (whereby the Gaussian was 
centered on the middle of the redshift bin and had 
$\sigma = ({\rm bin-width}/2)/3$,  
and we then use the Limber (1954) equation to transform to the 
angular correlation function.   We follow the formalism of 
the Tinker et al. (2005) n$_{\rm g}$ matched method to obtain the galaxy 
correlation function.   We find 
that our large scale clustering is in good agreement 
with the power law for the $z=1$ samples, however it is likely that 
the power-law should be steeper for such massive galaxies 
at high redshift, and the small scale clustering is at about the 
right amplitude.    We can therefore 
confidently say that the abundance matching reproduces the ball park 
correlation functions for galaxies, at a similar level to what is stated in 
Conroy \& Wechsler (2009).

\begin{figure}
 \vbox to 120mm{
\includegraphics[angle=0, width=90mm]{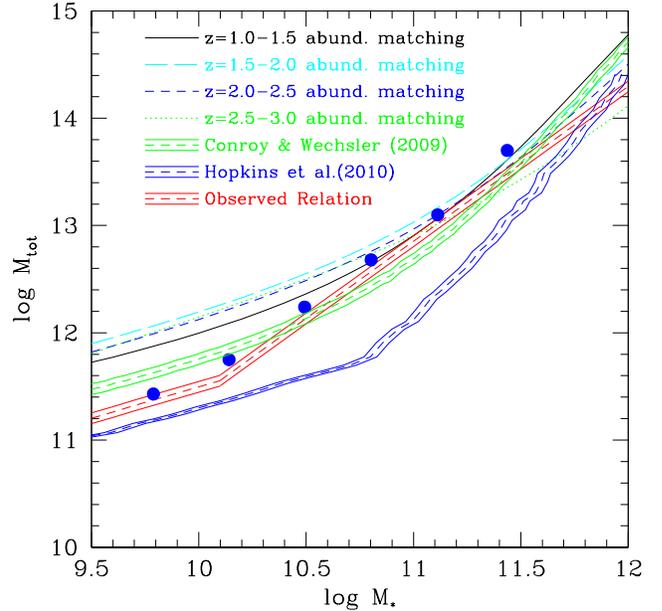}
 \caption{Figure showing the relationship between the stellar mass
and halo mass. The solid blue round points show this relation derived through
stellar mass measurements and measurements from kinematics (Twite
et al. 2014).  The other lines are from our halo abundance matching
fitting, while
the dashed line with two solid lines around it are from abundance matching
from Hopkins et al. (2010), and Conroy \& Weschler (2009).  }
} \label{sample-figure}
\end{figure}

We show the comparison between halo masses derived in this way and the stellar 
masses at the same limit in Figure~7 from redshifts from z = 1 to z = 3 using
two different methods. 
There is clearly little evolution in the halo to stellar mass ratio as a 
function of redshifts within our masses of interest, which is also what we find when we investigate this relationship 
using the observable relations from internal motions.  Figure~7 shows the 
relationship between the stellar and halo masses for galaxies up to $z = 3$ 
calculated in two different ways.  

\begin{figure*}
 \vbox to 120mm{
\includegraphics[angle=0, width=180mm]{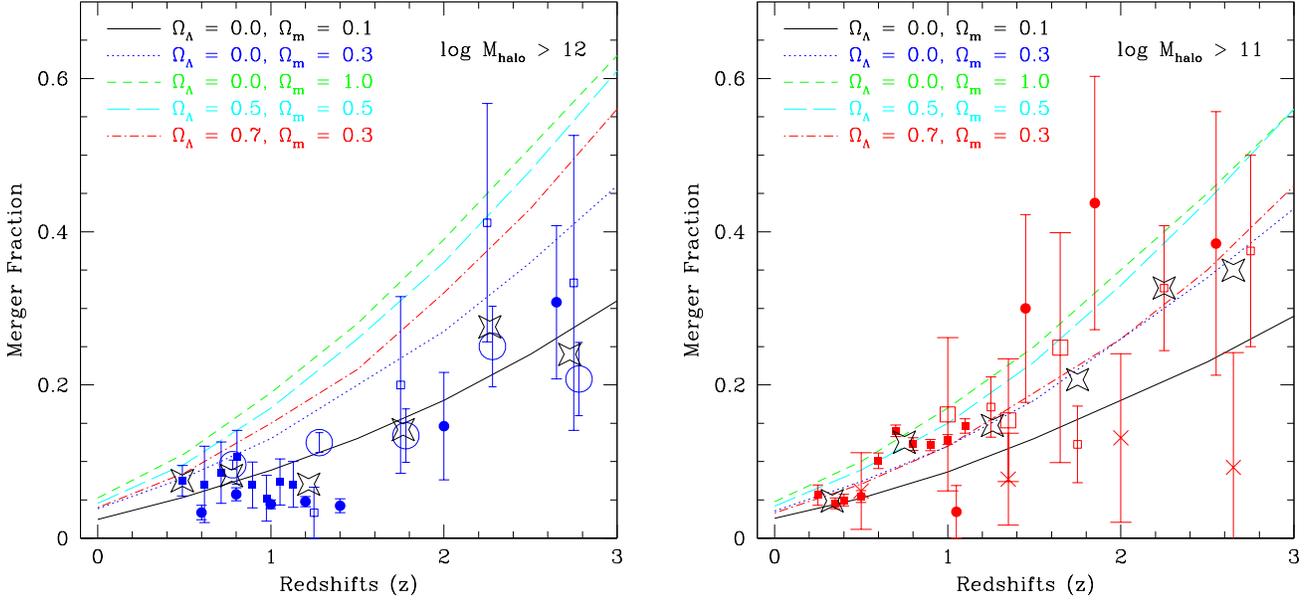}
 \caption{The halo merger histories compared with data for a) halos with masses \mhalo $> 10^{12}$ \solm and b) halos with masses \mhalo $> 10^{11}$ \solm.  The various lines show how the merger history varies with redshifts for halos of this given mass, and using the cosmologies listed in the upper left.
The solid symbols with errors
bars are data on the merger fraction evolution taken from Conselice et al. (2009) for those at $z < 1.5$, merger fractions from pairs from the 
Ultra Deep Survey (circles with inner error bars) and from
Bluck et al. (2009) for $z > 1.5$ in panel a)  -- all selected to have M$_{*} > 10^{11}$ \solm.  For panel b) the solid symbols with errors
bars are taken from Conselice et al. (2009) for those at $z < 1.5$ and from
Conselice et al. (2008) for $z > 1.5$, all selected to have 
M$_{*} > 10^{10}$ \solm.     For both the open symbols are from asymmetries 
from Mortlock et al. (2013) using CANDELS data.  The average error
weighted values of
the merger fractions from the various surveys are shown as open black stars. }
} \label{sample-figure}
\end{figure*}

There is some disagreement at the lower mass range, however this is 
due to the fact that there is unlikely a 1:1 galaxy - halo ratio at these
masses.  Hence we get an overestimate of the halo masses for these
systems. This is expected to some degree within this formalism, and this
effect has been seen before by e.g., Conroy \& Wechsler (2009).
 Note that we only use Figure~7 in the 
stellar mass range where our galaxies are found (log M$_{*} = 10-11.5$), and
do not consider these relations valid at high masses where the halo occupation
is very large and where this relation breaks down.

Shown in Figure~7 is the kinematic relation found by Twite et al.
(2014) between the stellar masses of galaxies and their total masses
as measured through kinematics.  The relation between these is given by:

        \begin{equation}
          \log(\mbox{M}_{\rm{tot}}) = (\alpha)\, \log(\mbox{M}_{*}) + (\beta),
          \label{eq:fit_all}
        \end{equation}

\noindent where the values of $\alpha$ and $\beta$ are a function of the stellar
mass.  We calculate (see also Twite et al. 2014) that for log M$_{*} < 9.9$, 
$\alpha = 0.58\pm0.03$, $\beta = 5.56\pm0.27$, while for systems with log 
M$_{*} > 9.9$,  $\alpha = 1.44\pm0.06$ and $\beta = -3.11\pm0.61$.  In Figure~7
we plot this relation along side the relation for the abundance matching.
In general there
is a correlation such that M$_{\rm halo}$/M$_{*} \sim$ 10, with a relatively
small scatter.  

This implies that we are able to match the halo mass to our
measured stellar masses for our galaxies in which we measure the stellar 
masses and merger histories from up to $z = 3$.    When we do these comparisons
we find that the log M$_{*}$ = 10 limit corresponds to log 
M$_{\rm halo}$ = 11.3 and log M$_{*}$ = 11 (see \S 5.2) corresponds 
to log M$_{\rm halo} = 12.7$.  We therefore use these limits for 
matching our observational data on mergers in galaxies
to those in halos.  Since we only use these two limits of stellar mass
we are almost always certain that these systems dominate their respective 
halos.

\subsection{Comparing Halo Mergers with Observed Galaxy Mergers}

With the caveats explained above, and using the halo mass to stellar mass 
comparisons in \S 5.1 we can now compare the halo merger histories to the observed galaxy merger fractions.

\subsubsection{Applying the Stellar Mass to Halo Mass Relation}

The key to performing the comparison between the observed mergers and the
predicted halo mergers is using the M$_{\rm halo}$ vs. M$_{*}$ relation
described in \S 5.1 to match
the halo mass with the stellar mass from our observations.  This must be
done carefully in two different ways. The first is that we must convert
the observed stellar mass of each galaxy to a halo mass.  However, we also
must be sure that the mass ratios used in the predicted halo mergers, in which
we use the value of $>1:4$, is the same as the mass ratio used in the 
selection of the observed mergers.
The issue is that if we use a $>$1:4 stellar mass ratio selection to
find mergers, using the relation
from eq. (15) this gives us a corresponding total mass ratio of $\sim$ 1:8.
Likewise, a halo mass ratio of 1:4 gives us a stellar mass ratio of
$\sim$ 1:2.6.  Therefore we have to ensure that any comparisons between
the halo merger fraction evolution, and the galaxy merger fraction are
done with the same underlying mass ratios.

For the mergers found through their high asymmetries, the merger ratio 
for galaxies
in total mass is $> 1:4$ based on simulations using dark matter (Conselice
2006) as well as through combined dark matter + baryon simulations (Lotz
et al. 2010).  However, this is not the case for the galaxies in pairs.  To
address this, we investigate the merger fractions for galaxies which
correspond to total mass mergers of $>$ 1:4.  This corresponds to a stellar
mass ratio of $>$ 1:2.6 using eq. (15).  We therefore calculate the 
stellar mass merger ratios using values of $>$ 1:2.6.  These merger
fractions with this ratio are therefore used on all the plots within 
this paper.    We determine the merger
fraction at this stellar mass ration by using the
 observations from Bluck et al. (2012).  We use this to correct other pair
fractions from Man et al. (2012) and  Lopez-Sanjuan et al. (2010), 
with the assumption that the relative
fractions change in the same way as in the Bluck et al. (2012) study. Regardless the vast majority of our comparisons are done using the CAS mergers where
the total mass ratios are already $>$1:4.

\subsubsection{Comparison of Observed and Predicted Mergers}

Using the information above we show in Figure~8 a direct comparison between 
the halo merger history predictions
which we have discussed throughout this paper, and the observed 
galaxy merger histories
based on our stellar mass selection.  
Figure~8 shows that there is a general agreement between the 
shape and normalization of the merger history for the predicted halos and the
observed galaxy merger histories.  To make a quantitative comparison we 
use the conversion factors from eq. (15), with its associated uncertainties
to determine the halo mass for the corresponding selection of stellar mass.
For log M$_{*} > 10$ this corresponds to a selection of log M$_{\rm halo} =
11.3\pm0.4$.  We then use the relations discussed in \S 3.1 to determine
the merger fraction for galaxies at this halo mass and their associated
uncertainties as a function of redshift.  As discussed in \S 3.1, there is very little variation
in the merger fraction for halos of different masses at a given redshift.
For example, at $z \sim 2.5$, the merger fraction with an uncertainty given
by the models and conversion uncertainty is 
$f_{\rm halo} = 0.35^{+0.01}_{-0.02}$.

Our relatively large observational errors on the merger fraction evolution 
cannot easily distinguish between these various models 
currently, an issue we discuss in more detail below in terms of future surveys 
that can improve this comparison with data.   We however carry out a full analysis
of the best fit model using using a reduced $\chi^{2}$ approach,  and using
the uncertainties calculated using the methods described above
across all redshifts. We do this
by matching the model with the corresponding halo mass to our stellar mass
derived galaxy merger measurements.  We then utilize the best fitting
power-laws to determine the $\chi^{2}$ of each fit.  The reduced
$\chi^{2}$ values range from 4.8 to 14.43, with the best fitting model
the concordance cosmology with $\Omega_{\Lambda} = 0.7$ and $\Omega_{\rm m}
= 0.3$.   The data also rules out that
the universe has a low matter density of \om $= 0.1$, although in terms of
the comparison there is a similarity between the merger history
for models with \om = 0.3, \ol = 0.7 and \om = 0.3, \ol = 0.  

We furthermore show in Figure~10 the variation of the halo merger
fraction as a function of $\Omega_{\Lambda}$ at $z = 2.5$ with the 
assumption that $\Omega_{\Lambda} + \Omega_{\rm m} = 1$ holds throughout.  
The best fit
relation between the merger fraction and the value of $\Omega_{\Lambda}$ is
given by:

\begin{equation}
f_{\rm halo} = \alpha \times \Omega_{\Lambda} + \beta = \alpha \times (1-\Omega_{\rm m}) + \beta
\end{equation}

\noindent Where the values for the fit for $\Omega_{\Lambda} < 0.55$ is
$\alpha = -0.05\pm0.01$ and $\beta = 0.41\pm0.01$.  For  
$\Omega_{\Lambda} > 0.55$ the best fit is given by
$\alpha = -0.24\pm0.02$ and $\beta = 0.52\pm0.01$.
The relation between the halo merger values and $\Omega_{\Lambda}$
is such that at $\Omega_{\Lambda} < 0.55$
there is very little variation between the value of $\Omega_{\Lambda}$ and
the merger fraction.  This relation becomes steeper for values
$\Omega_{\Lambda} > 0.6$, making it a more sensitive measurement of
$\Omega_{\Lambda}$ and $\Omega_{\rm m}$.

Using our best measured merger fraction of $f_{\rm m} = 0.31 \pm 0.07$ at 
$z = 2.5$, we find that this formally leads to a measured 
$\Omega_{\Lambda} = 0.84^{+0.16}_{-0.17}$, which is very uncertain compared to
other leading methods of finding $\Omega_{\Lambda}$, but still demonstrates
consistency with previous work, and that there is at least a broad agreement
between cosmology and galaxy formation as seen through mergers.  The error
bars on this measurement come from the uncertainty in the theoretical fit and 
the uncertainty in the measured merger fraction from Mortlock et al. (2013)
based on CANDELS data and incorporating stellar mass and redshift uncertainties.

\begin{figure}
 \vbox to 120mm{
\includegraphics[angle=0, width=90mm]{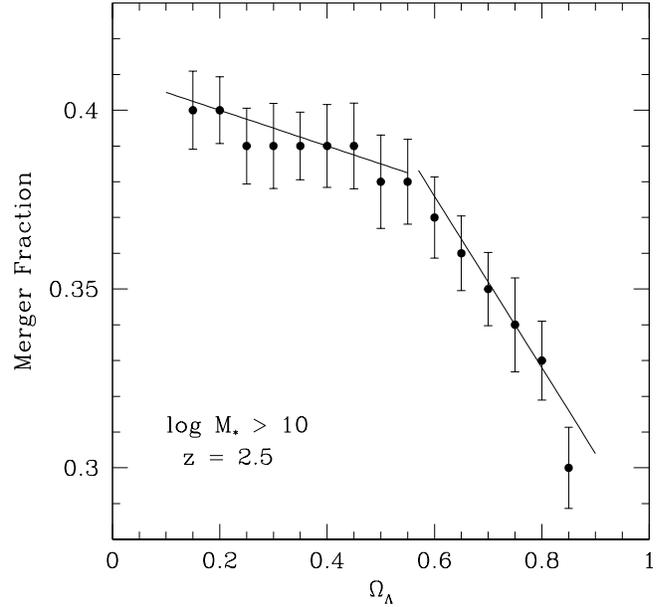}
 \caption{The relationship between the value of $\Omega_{\Lambda}$ and
the resulting merger fraction for systems with stellar masses
$> 10^{10}$ \solm at z = 2.5.  This is for the condition such
that $\Omega_{\Lambda} + \Omega_{\rm m} = 1$.  The solid lines are
the best fitted relationship for this relationship as described in
the text.  }
} \label{sample-figure}
\end{figure}

We can also now compare the merger history to the predicted halo mergers with differing
values of the $\sigma_{8}$ parameter.  Figure~9 shows this comparison of our 
merger histories from observed galaxies in comparison to the halo mergers for 
differing values of $\sigma_{8}$.  
What we find is a relatively high value of the merger fraction in comparison
to the halo merger histories.  This suggests that the value of $\sigma_{8}$,
as derived from our observations would be low, with the value of $\sigma_{8}$
= 0.7 in best agreement with our observations.  In the next sub-sections we 
discuss how to use observed mergers as a competitive tool to constrain 
cosmology.

\begin{figure}
 \vbox to 160mm{
\includegraphics[angle=0, width=90mm]{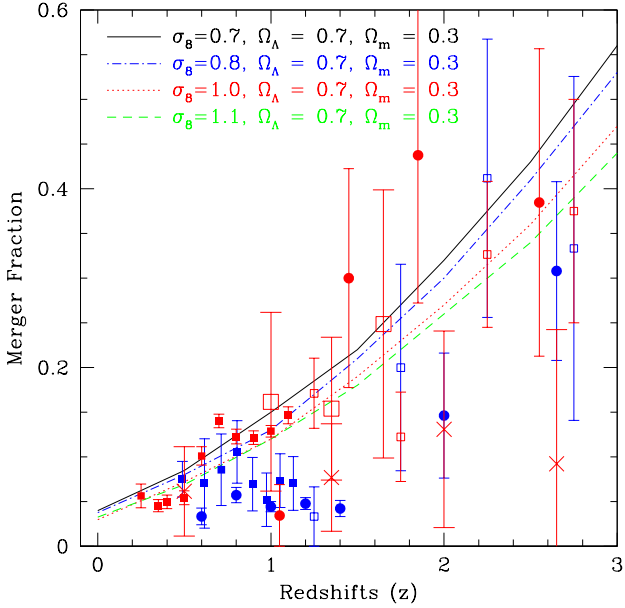}
 \caption{The evolution of the halo merger fraction for halos with
\mhalo $> 10^{12}$ \solm as a function of the value of $\sigma_{8}$ 
as in Figure~3 except that here we show a comparison to data.
The range shown is from $\sigma_{8}$ = 0.7 to $\sigma_{8} = 1.1$,
 The points shown here are
actual measures of the merger fraction for galaxies as a function of
redshift.  The red points are for  M$_{*} > 10^{10}$ \solm
galaxies, including results from Conselice et al. (2003) at $z > 1$ in
the HDF (solid circles); Conselice et al. (2009) (solid boxes at
$z < 1.2$); and Mortlock et al. (2013) (small open boxes at $z > 1$).   Also 
shown are pair merger fractions at separations of $< 30$ kpc: (Man et al.
2012; crosses at $z > 1$); Lopez-Sanjuan et al. (2010) (large open boxes).
The blue points show the merger history for M$_{*} > 10^{11}$ \solm
systems including results from Conselice
et al. (2009) (solid boxes at $z < 1.2$; Bluck et al. (2009, 2012)
(solid circles at $z > 0.5$); and Mortlock et al. (2013) (open boxes
at $z > 1$).   }
} \label{sample-figure}
\end{figure}
\vspace{10cm}

\subsection{Cosmologies with an Evolving Equation of State}

In the previous sections we investigate the merger history and how it
compares to dark matter halo mergers for a variety of cosmologies.  However,
it is clear that the concordance cosmology is the most likely cosmology with the
large amount of supporting observations in the past few decades.  What is
not known for certain is the role of an evolving form of the dark energy, and
how this may affect the galaxy formation process.  In this section we investigate
how the merger history for halos changes for a variety of evolving equations
of states of the universe itself. 

In fact, one of the major goals in cosmology is to constrain the equation of
state of the universe, which relates the pressure (P) and density
($\rho$) of the dark energy, or $\omega = P/\rho$.   The value of
$\omega = -1$ is for a cosmological constant.   The best current estimates
of $\omega$ give values $\omega = -1.08\pm0.08$ (Anderson et al. 2012).
The dark energy equation of state can also be written as 
$\omega(z) = \omega_{0} + \omega_{a} (1-a)$, although the measurements
of the parameters $\omega_0$ and $\omega_{a}$ are not well
constrained, with the latest measurements giving $\omega_{0} = -0.905\pm0.196$
and $\omega_{a} = -0.98\pm1.09$ (Sullivan et al. 2011).

In this section we investigate how the merger history varies with different
values of the equation of state of the universe. These merger predictions
are from the GALACTICUS models of Benson et al. (2012) using the equations
from Percival (2005).
First we evaluate how the merger fraction varies for halos with masses
M$_{\rm halo} > 10^{11}$ \solm for values of $\omega$ between $\omega = -0.33$
to $\omega = -1.2$ (Figure~11), with $\omega_{\rm a}$ = 0 for both.  
We also investigate this at two different
halo masses, those with log M$_{\rm halo} > 11$ and for 
log M$_{\rm halo} > 12$.
This shows a couple of interesting features regarding how the values of
$\omega$ influence the halo merger history.  We make the note that, as
clearly shown in Figure~11, it will be very difficult to utilize this
type of figure to constrain the value of $\omega$ with galaxy merger
histories given that there is not a large range of values 
of the halo merger fraction at different redshifts.

However, there are a few trends we can see. The first is that higher
negative values of the equation of state parameter, $\omega$, gives
a higher merger fraction down to $z = 1.4$.  However, at redshifts
lower than this the models become nearly
indistinguishable from one another.  Furthermore, as in the basic merger
history, the higher mass systems have higher merger fractions.

These changes in the different $\omega$ merger histories are well fit
by a power-law of the form $(1+z)^{m}$ up to $z = 3$.  We can use this
to investigate the likelihood of being able to detect a difference
in the merger history which may help constrain cosmology.  We find
that the difference between the effective merger histories is 
roughly $\delta f_{\rm halo} \sim 0.005$ from values of $\omega = -0.9$ 
to $\omega = 1.1$.   We
discuss how this could be useful in cosmological investigations in \S 6.

We also investigate how the merger history varies with an evolving form
of the equation of state, such that 

\begin{equation}
\omega(z) = \omega_{1} + \omega_{\rm a}(1-a)
\end{equation}

\noindent where there is currently very little constraint based on observations
of supernova (Sullivan et al. 2011).  We however find that the merger 
histories for all reasonable combinations of $\omega_{1}$ and 
$\omega_{\rm a}$ are very close to one another in merger space (Figure~12), 
such that the differences in $\delta f_{m} \sim 0.005$, again close to the 
accuracy needed to probe similar realistic differences  for different 
constant $\omega$ values.  This however becomes larger at higher redshifts
such that the difference between different values for varying equations 
of state is highest.   In the future observing the merger fraction at 
$z \sim 6$ will be a major way to make this distinction using the ELTs 
and the JWST.

\begin{figure*}
 \vbox to 120mm{
\includegraphics[angle=0, width=180mm]{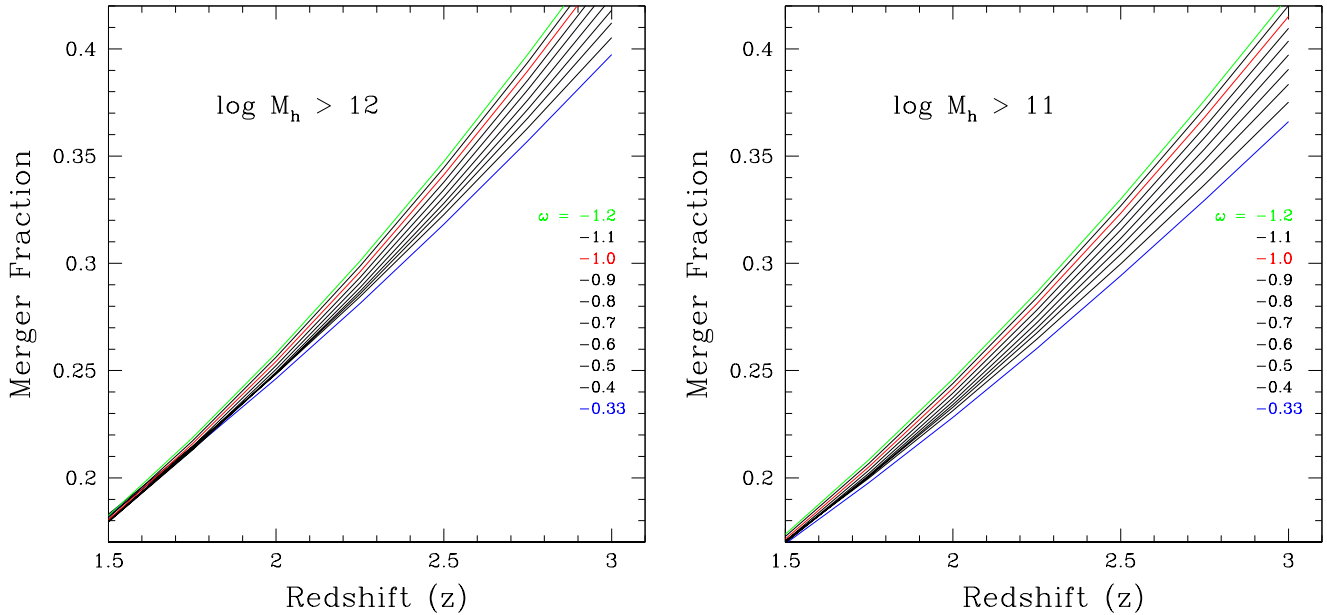}
 \caption{The change in the halo merger fraction as a function of
redshift for differing values of the cosmological constant, as
given in the equation of state for the dark energy, $\omega$ (\S 5.3).
The left panel shows the merger history for halos of masses
M$_{\rm halo} > 12$, while the right is for halos of mass
M$_{\rm halo} > 11$.  The green line and blue lines show the
range of our probe from $\omega = -1.2$ down to $\omega = -0.33$.
The red line shows the values for the fiducial $\omega = -1$. }
\vspace{0cm}
} \label{sample-figure}
\end{figure*}

\begin{figure}
 \vbox to 120mm{
\includegraphics[angle=0, width=90mm]{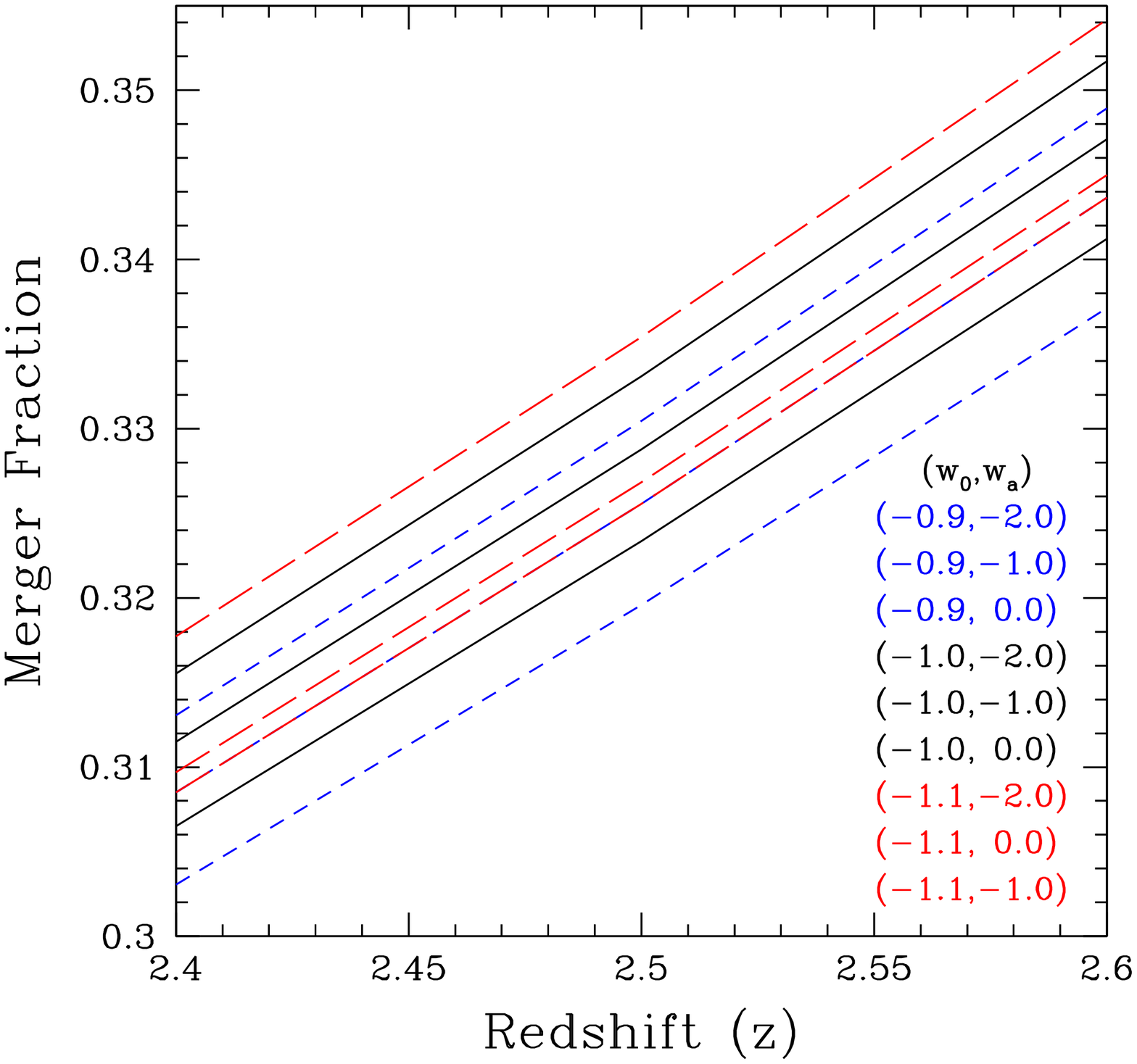}
 \caption{The relationship between the merger history as a function of 
redshift and different values of $\omega_{0}$ and $\omega_{1}$ for
an evolving form of the dark energy equation of state (see \S 5.3).  }
} \label{sample-figure}
\end{figure}

\subsection{Effects of Cosmology on Observables}

One of the issues that we have to address when comparing galaxy observables,
such as the merger history to models of varying cosmologies, is that the
observables we use depends on the cosmology assumed when calculating
derived values such as the stellar mass or merger rate.  For example, the
mergers we find are based on measuring a stellar mass and the value of this
stellar mass can change depending on the cosmology used to calculate it. 

Other features which change with cosmology are the physical separation between
two galaxies, and whether or not this is a physical pair or not according
to our definition based on the angular size distance. We therefore investigate
in this section how the observed merger fractions and pairs can depend upon the
cosmology assumed.  We investigate these differences in terms of the concordance cosmology.

\begin{table}
\centering
\begin{tabular}{c c c r}
\hline \hline
$\Omega_{\rm \Lambda}$ & $\Omega_{\rm m}$ &  Ang. Size Dist Ratio & $\Delta M$  \\
\hline
0.7 & 0.3 & 1.0 & 0.00  \\ 
0.0 & 0.1 & 1.4 & 0.67 \\
0.0 & 0.3 & 1.2 & 0.17 \\
0.0 & 1.0 & 0.9 & -0.10 \\ 
0.5 & 0.5 & 1.1 & 0.09 \\
\hline
\end{tabular}
\caption{How galaxy observables change as a function of cosmology, most notably the
angular size distance and the luminosity distance, which has a direct effect
on the measured stellar masses.  Shown here is the ratio of the angular size
distance, defined as the angular size distance at the given cosmology divided
by the angular size distance in the concordance cosmology.  Likewise the
stellar mass differences between these two cosmologies is shown.} 
\end{table}

We show in Table~6 the values of the difference in derived stellar mass 
$\Delta M_{*}$ within the various cosmologies in reference to the coordinate cosmology.  The biggest
change is that at a given stellar mass selection, the galaxies examined
are up to $\delta M_{*} = 0.67$ more massive depending on the cosmology examined.    
Likewise we find that the angular
scale can vary by 40\%, but mostly around 20\% for cosmologies different
from the concordance cosmology.

We investigate how the merger fraction changes when the measurements of stellar masses
and galaxy separations are affected by different cosmologies. Note that
the cosmology has more of an effect on the measurements for mergers which are
found within pairs of galaxies.  For morphological mergers, there is no angular
separation to worry about and thus it is only the change in the distance
modulus with cosmology that affects the choice of which bin a galaxy belong in.
 
For this reason we  utilize merger fractions, f(M$_{*}$) which are a pure observable, 
as a function of
stellar mass, which is as noted a function of the cosmology.  We
could use merger rates, but the time-scales for these merger rates depends to
some degree on cosmology, and on the nature of the dark matter. By using merger
fractions we are able to side-step this issue to some degree as it is a purely
observational quantity.  

To understand this issue we examine how the merger fraction varies with pairs 
when changing the cosmological parameters, which results in the  different
selections of underlying halos.   We find that the observed
merger fractions increase at larger separations, as long noted in pair studies of
galaxies, and increases for higher mass selections at $z > 2$. 

The result of this is that the cosmology with a low matter density,
those with $\Omega_{\rm m} = 0.1$, $\Omega_{\Lambda} = 0$ will sample
at a given angular separation, and a given flux cut lower mass galaxies
and smaller separations as both the luminosity distance and angular size
distance is larger in this cosmology than in the concordance
one.  To account for this cosmology we would have to examine intrinsically
fainter galaxies and look for pairs which have a smaller angular separation on
the sky to match the intrinsic conditions used to find mergers in the
coordinate cosmology.  Both of these effects will decrease the merger
fraction measured.  If we take the $\Delta M = 0.67$ mass difference
and the factor of 1.4 larger angular size separation we would
measure the merger fraction as $\delta f_{\rm m} \sim 0.05$ lower
than what is plotted on Figure~8.  This is within the errors of the
merger fractions themselves, but would decrease the measurement towards
the model values further than the observed.

\begin{figure}
 \vbox to 120mm{
\includegraphics[angle=0, width=90mm]{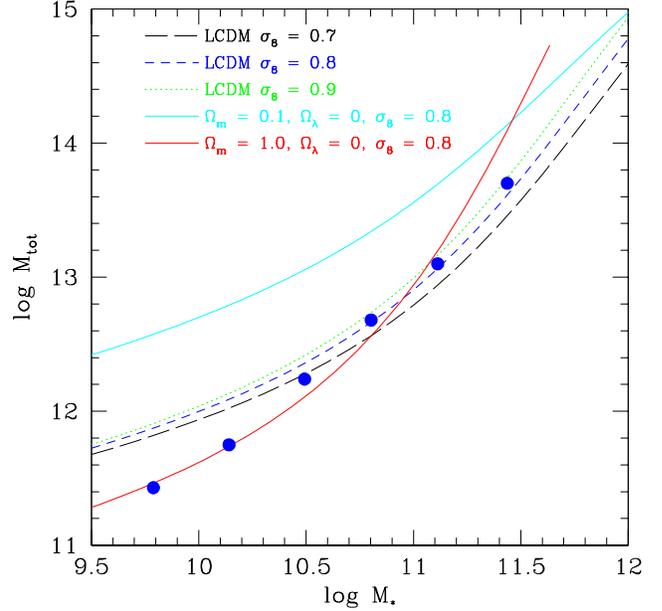}
 \caption{The stellar mass and total mass relation using the
abundance matching methodology described in \S 5.1, but using
different cosmologies within the calculation for the relations
at $z = 1 - 1.5$.    The dashed lines are for the concordance
cosmology but using different values of $\sigma_{8}$.  The
solid lines are for $\Omega_{\rm} = 1.0$ (red) and
$\Omega_{\rm m} = 0.1$ (cyan), with both at $\Omega_{\Lambda}$ = 0.
The blue points are the kinematic relations for the concordance cosmology
described in \S 5.1.  }
} \label{sample-figure}
\end{figure}

Likewise, universe's with a high mass density, i.e., with
$\Omega_{\rm m} = 1$ have smaller angular size and luminosity
distances.  Therefore at a given flux, and at a given angular
size measure on the sky, the galaxies sampled within this
cosmology will be of lower mass and larger intrinsic separations.  This gives the
opposite effect from above, and to measure the correct merger fraction would
require going to brighter systems and at larger angular separations.  These
both would result in a higher merger fraction based on measurements of
how the observed merger fraction changes at fainter limits and smaller
separations.  We calculate for this cosmology that the true comparison
merger fraction would be  $\delta f_{\rm m} \sim 0.05$ higher,
 which would approach the higher values of the models, but would still
be far from matching better than the concordance cosmology.  More detailed
comparisons will therefore have to be carefully done and all merger
fractions measured in the cosmology being compared with.

There is also the issue that our mapping of stellar and halo masses will 
depend on the cosmology assumed in the calculation.  We show these 
different stellar vs.
halo masses using different cosmologies in Figure~13.  For the 
concordance cosmologies with differing values of $\sigma_{8}$ 
from 0.7 to 0.9 there is a variation of
the total mass for a halo mass at log M$_{*} \sim 11$ is 0.3 dex.  This would
thereafter lead to a change in the derived merger fraction by at 
most 10\%.  As can also
be seen, there is a large change in the measurement of the mass relations for
the cosmologies with $\Omega_{\rm m} = 0.1$ and $\Omega_{\rm m} = 1$.  These
differences lead to a factor of ten difference in the halo mass at a given
stellar mass  from the concordance cosmology at 
masses around log M$_{*} \sim 11$ in the most extreme examples. However, as
the halo merger fraction dependence on mass is quite shallow, this only
leads to an addition small uncertainty in the halo merger fraction of 
f$_{\rm halo} \sim 0.025$.

When better merger fractions are available they can be
measured in different cosmologies and thereby produce a more detail comparison
between models and data.   In practice this will consist of calculating stellar masses
and galaxy merger fractions at the different cosmologies under study.  This  
should not be complicated to do, only time intensive.  We finally note that the use of 
morphology and structure to find galaxies merging
is potentially a more powerful approach than those from pairs of galaxies.  The
reason is there is less effects from cosmology on the measures due to only the
stellar mass selection being affected since there is no selection by angular
separation of two galaxies.

\section{Discussion}

\subsection{Mergers as a Cosmological Tool}

In this subsection we investigate using the merger history as a cosmological
tool. We preface this by stating that mergers may never be as good of a 
method as some to measure cosmological
properties. The reason that methods that use standard candles such as Type
Ia supernova are so successful is that fundamentally there are only two  
observations - the flux and redshift, which can then be compared directly 
to cosmological predictions.
However, our method does allow us to measure cosmology by the effect it 
has on objects within the universe,  rather than its effects of the 
expansion which most methods use.

The use of the galaxy merger history as a competitive cosmological tool will 
require that we know more about the properties of galaxies, namely accurate 
measurements
of galaxy masses and the ability to understand the merger history without
significant systematic biases.  Although the merger history has some
agreement using different methods such using galaxy pairs and through
morphology (e.g., Conselice et al. 2009; Lotz et al. 2011), there still
remains work to be done to determine whether or not we can actually measure
the merger fraction or merger rate to within 1\% up to $z \sim 3$.  
This means we will have to measure
the merger history accurately over all environments, such that the effects
of cosmic variance produces less than a 1\% systematic error on the measured
merger values.

Even if we can measure the merger history accurately without significant
systematic errors, there is still the issue of shot noise within these
measurements, requiring larger area surveys than what we currently
have.   We can investigate
how good our merger fraction measurements will need to be to measure 
the history of mergers to distinguish between cosmological models.  

To measure accurately the differences in the cosmological models presented 
in this paper requires that the accuracy in measuring merger fractions for
galaxies at a given mass is $\delta f_{\rm m} \sim 0.01$.
To distinguish between $\delta f_{\rm m} \sim 0.01$, for a merger fraction of
$f_{\rm m} \sim 0.3$, i.e., a 30\% fraction of the population undergoing
mergers, which we see at $z > 2$ at high stellar mass, we need to measure
accurately the merger fraction over 4335 arcmin$^{2}$ or 
1.2 deg$^{2}$.  Surveys this large do now exist, such as the CFHT legacy survey and
the UKIDSS Ultra-Deep Survey (UDS), and the next generation
of photometric and spectroscopic redshifts within these should make an attempt at measuring
merger fractions by using pairs.  However, the use of morphology for
the measurement of merger fractions will be difficult given that only a very
small fraction of this area has been imaged at high enough resolution in the 
near-infrared in the CANDELS survey (e.g., Grogin et al. 2011; Koekemoer et al.
2011).

We can further investigate the accuracy we need to get a 3 $\sigma$ accuracy on 
statistical measures of merger fractions.  If we want to obtain a 3 $\sigma$  measurement 
at $\delta f_{\rm m} \sim 0.01$ we will need
a larger area of 8.1 deg$^{2}$.  This is much larger than any deep surveys,
yet in the future, those with Euclid or WFIRST will be able to measure
the structure/morphologies over this area.

Attempting to use the merger history for measurements of different values of
the equation of state parameter, $\omega$, and for an evolving form of $\omega(z)$,
require a precision about twice as good.  As briefly discussed in \S 5.3, to measure
different merger histories that can constrain between small values of $\omega$
to within $\pm 0.1$ around $\omega = -1$, or to constrain better than current
estimates of $\omega_{1}$ and $\omega_{\rm a}$ requires that we measure the merger
fraction to within $f_{\rm halo} \sim 0.005$ or better.  This will require deep surveys
of at least 2.5 deg$^{2}$ in area, which is currently possible by combining
existing surveys.  
To obtain 3 $\sigma$ statistical errors will require an area of $\sim 17$ deg$^{2}$.
This also assumes that other errors, such as systematics with redshifts, stellar
masses, etc. are small, which they are not.  Along side obtaining deeper and wider
surveys we will also need to be more certain that accurate stellar masses
and photometric/spectroscopic redshifts can be measured.  As these improve
to measure galaxy evolution so will our ability to use galaxies as a cosmological
tool.

\subsection{Implications}

There are a few important implications within our results, for cosmology,
structure assembly, galaxy formation, as well as for our ability to measure
the merger history.   We briefly detail these here, although a further analysis
of these will be addressed in later papers. The first issue is that
by examining the dark matter assembly itself, rather than relying on the baryonic
physics behind semi-analytical simulations we get a much better agreement
with the merger data, making reasonable assumptions about the relation 
between the
observed stellar mass of galaxies and their total halo mass from known 
scaling relations.    This is an important point to make, as it reflects the
well known problem of matching the abundances of galaxies with their
total halo mass functions.  It is clear that the galaxy mass function has a
shape which differs significantly from the predicted halo mass function and
processes such as feedback, photoionization, conduction, gas cooling among
other baryonic processes must be implemented to correctly model the 
observed galaxy mass function (e.g., Benson et al. 2003).

However, for whatever reason the implementation of the baryonic physics in
semi-analytical models does not accurately reproduce the merger fraction and
rate history of galaxies.  This implies that the baryonic physics 
implemented in these models is missing ingredients necessary to
reproduce the observed history of galaxy assembly, or is implementing
them in an incorrect way.  This is perhaps 
also reflected in the problem of semi-analytical models in predicting the
number of massive galaxies at high redshift (e.g., Conselice et al. 2007; Guo
et al. 2011).
While the Warm Dark Matter models do better, this may reflect less structure
on smaller scales, and thus more likely to produce similar mass mergers.

We also demonstrate that the merger history depends on cosmology such that
cosmologies with higher matter densities will contain a high merger fraction
and rate, and therefore that
there will be more massive halos in these cosmologies found in the nearby 
universe.  We furthermore show, in a limited way, that there is also a 
good agreement between the merger history and the generally accepted 
cosmological model, such that \om = 0.3 and \ol = 0.7 is the best fit 
model based on a wide range of possible cosmologies.  However, it is 
worth pointing out that
this constraint is not very useful in itself given that the standard 
model of cosmology is generally accepted. To be a useful indicator 
for cosmology, this method 
would have to be able to distinguish suitable differences in the merger
history for the limited range of cosmologies that are now predicted in
CMBR experiments such as WMAP and Planck (e.g., Komatsu et al. 2011; 
Ade et al. 2013).  However, it is reassuring that the agreement we find 
demonstrates a consistency between the dominant and generally accepted 
cosmology, and how the dark matter on the level of galaxies is assembling
through mergers.

It is possible to distinguish between various cosmologies with various
\om and \ol values, however we conclude that the accuracy in the merger
fraction needed to do this is very high, on the order of 1\% to distinguish
between various cosmologies with similar values.  This will require surveys
from 1.2 deg$^{2}$ in area up to 20 deg$^{2}$, but the real challenge will
be in making sure that all systematics are accounted for when carrying out 
this comparison.  One of these challenges is understanding the merger
time-scale, that is how long an asymmetric galaxy, or a galaxy seen in a pair,
takes to merge.  This is essential when comparing with any predictions.  
The merger time-scale is estimated to be 0.4 Gyr in this paper, which is
what is found empirically by observing the change in the observed merger
fraction (Conselice 2009), as well as an average over simulations (e.g.,
Conselice 2006; Lotz et al. 2010).  However, the merger time-scale does
depend upon the galaxy gas mass fraction (e.g., Lotz et al. 2010) and 
other properties such as viewing angle, bulge to disk ratio and orbital
type.  While the average of such properties may be calculable from models,
it remains to be determined whether time-scales seen at high redshift are
similar to those predicted in models.  Future detailed simulations with
for example {\it Eagle} and {\it Illustris} will help reveal time-scales for these
mergers which will eliminate another key systematic to create these
comparisons.

To avoid the time-scale issue completely the merger fraction comparison
could be carried out using galaxies within close pairs -- i.e., systems
which are just about to merge.  Simulations can predict this often as
well as they can an actual merger, and observationally, these pairs
are not as ambiguous at times as deciding if a galaxy is an active merger
or not.  Likely, a combination of pre-merger close pairs and post-merger
morphological signatures are best used together as a check on systematics of
both methods.

We also note that the agreements with the theory for some of our merger 
histories is not perfect.  The most obvious case of this is the comparison 
between the
halo merger history and galaxy mergers selected with stellar 
masses M$_{*} > 10^{11}$ \solm
(Figure~8).  Here we can see that at $z < 1$ the galaxy merger 
history is significantly less than that predicted based on the 
halo mergers.  This is likely
an indication that we are simply not detecting all the mergers 
for this population of very massive galaxies at this epoch. The 
reason for this is that many of these massive galaxies are fairly 
red at this epoch, and any mergers would be `dry' and thus not be 
detectable within the CAS systems, from which these $z < 1$
values are measured (e.g., Hernandez-Toledo et al. 2006).  What is needed is
a full analysis of the merger history at this mass range using galaxies
in kinematic pairs, as well as through clustering analyses to determine the
actual merger rate.  

It is likely that the ultimate measurement of the merger history will be
carried out through a mixture of galaxies in pairs, and those measured through
their high asymmetries.  Typically the morphological approach is useful at higher
redshifts, while the pair method is more suitable for lower redshift galaxies,
especially at $z < 1$ where there are significant investments in spectroscopy
within deep fields.  

\section{Summary}

This paper explores the idea of using galaxy mergers as a probe of cosmology. 
While the idea that galaxy merging can be used to probe the properties of
the universe is not new (e.g., Carlberg 1991), the use of it as a cosmological
probe has never been fully characterized, or explored.  In this paper we have
shown that using a variety of assumptions regarding how the halo merger history can
be matched to some degree with the galaxy merger history, the
current observations of the galaxy merger history are in relative agreement
with a concordance cosmological model.  

Our other major findings are:

\noindent 1. The halo merger history varies as a function of halo mass, such that 
systems with larger halo masses have higher merger fraction at all
redshifts.  This is in direct contrast to the observations of galaxy mergers
whereby the most massive systems (as measured in stars) have a higher merger
fraction at $z > 2$, but tend to show little merging at later times.

\noindent 2. Semi-analytical models based on CDM 
from the Millennium simulation under-predict the galaxy merger history, but
that those using Warm Dark Matter with a particle temperature of
$\sim 1$ keV do a better job in predicting the merger
history than a CDM semi-analytical model.

\noindent 3. The merger history of halos varies significantly with cosmology. 
The merger fraction and history is lower for lower matter density $\Omega_{\rm m}$
cosmologies, and highest for cosmologies where $\Omega_{\rm m} = 1$.  The difference
between these is large, around $\delta f_{\rm halo} \sim 0.25$ at $z \sim 3$ and
easily distinguishable with current measurements. 

\noindent 4. We show that it is possible to compare halos to galaxies through the use of
halo mass to stellar mass calibrations, and thus to compare directly halo mergers
to galaxy mergers.  When comparing the available
data to these models the best fit is a concordance cosmology with $\Omega_{\Lambda}
= 0.7$ and $\Omega_{\rm m} = 0.3$.  We also show how the merger fraction varies
as a function of $\Omega_{\Lambda}$ at a single redshift, $z = 2.5$, and how
this can be further used to calculate the best fit value of 
$\Omega_{\Lambda} = 0.84^{+0.16}_{-0.17}$

\noindent 5. The halo merger history is also strongly dependent on the value of
$\sigma_{8}$, such that lower values of $\sigma_{8}$ give a higher merger fraction.

\noindent 6. We also examine how the merger history changes for differing values of
the dark energy equation of state
$\omega$, and how a varying $\omega(z)$ changes the calculated halo merger history.
We find that accurate merger fractions on the level of $\delta f \sim 0.005$ are
required to distinguish between competing models.  The difference between the predictions
for the merger histories is highest at the highest redshifts and in the future JWST and
the E-ELT can provide these measurements.

The use of the galaxy merger history to probe cosmology in
a competitive way with other techniques such as supernova, baryonic acoustic
oscillations, and CMBR work will require several improvements.  The first is that
 we must be able to match better the  stellar masses of
galaxies and their halo masses, as well as have a firmer idea of the
mass ranges that produce merging and the time-scale for halo mergers,
and how these relate to the time-scale for galaxy mergers.  While halo
occupation is one way to do this, and can produce successfully matched
merger histories, we are able to derive similar results using a calibration
based on the kinematics of galaxies.

Overall, we find that the accuracy of merger fractions would have to exceed
$\delta f_{\rm m} \sim 0.01$ to be able to differentiate between current
uncertainties in the value of cosmological parameters.  As explained
in this paper this will required surveys on the sky of area  up to
several 10 deg$^{2}$.  These types of surveys must also have accurate stellar
masses and reliable redshifts with uncertainties, along with the merger
uncertainties, which do not add up to a merger fraction errors that are larger
than 1\%.  This will be difficult, but by combining results at various 
redshifts, and various masses, some of these limits on uncertainties 
could be relaxed a bit and systematics
better understood.  Future surveys such as Euclid and LSST will be ideal for
carrying out this type of analysis and has a potential to be competitive with
other techniques such as those listed above.  

We thank Mike Santos, Aisyah Sahdan and Jack Johnson for their early 
contributions to this work, and Ed Copeland for illuminating discussions.  
CJC, AM and AFLB acknowledge support from the STFC and the Leverhulme Trust.  
D.P. acknowledged the support 
of an Australian Post-graduate Award and an Endeavour Research 
Fellowship. A.F.L.B. acknowledges support from NSERC (National Science and 
Engineering Research Council) of Canada.

\appendix

\label{lastpage}

\end{document}